\documentclass[11pt,letterpaper,twoside,notitlepage]{amsart}

\let\proof\undefined
\let\qed\undefined

\usepackage{amsmath}%
\usepackage{amsfonts}%
\usepackage{amssymb}%
\usepackage{graphicx}
\usepackage[margin=10pt,font={small,it},labelfont=bf]{caption}
\usepackage{nccthm}
\usepackage{fancyhdr}
\usepackage{fancybox}
\usepackage{tabls}
\usepackage[left=2.75cm,top=2cm,right=2cm,bottom=2cm,includehead]{geometry}

\numberwithin{equation}{section}

\begin{document}

\begin{titlepage}
\begin{flushright}
      arXiv:0705.0545v1\\
      ITP--UH-10/07\\
\end{flushright}
\vspace{20mm}
\begin{center}
{\LARGE\bf  On Verlinde-Like Formulas in $\mathbf{c_{p,1}}$ Logarithmic Conformal Field Theories}\\
\vspace{5mm}
{\sc Michael Flohr} and {\sc Holger Knuth}\\
\vspace{5mm}
{\small May 4, 2007}
\end{center}
\vspace{10mm}
\begin{center}
{\bf Abstract}\\
\vspace{3mm}
\parbox{130mm}{
\noindent Two different approaches to calculate the fusion rules of the $c_{p,1}$ series of logarithmic conformal field theories are discussed. Both are based on the modular transformation properties of a basis of chiral vacuum torus amplitudes, which contains the characters of the irreducible representations. \\ One of these is an extension, which we develop here for a non-semisimple generalisation of the Verlinde formula introduced by Fuchs et al., to include fusion products with indecomposable representations. The other uses the Verlinde formula in its usual form and gets the fusion coefficients in the  limit, in which the basis of torus amplitudes degenerates to the linear dependent set of characters of irreducible and indecomposable representations. We discuss the effects, which this linear dependence has on any result for fusion rules, which are calculated from these character's modular transformation properties. \\
We show that the two presented methods are equivalent. Furthermore we calculate 
explicit BPZ-like expressions for the resulting fusion rules for all $p$ larger than $2$.
}
\end{center}
\vfill
\end{titlepage}

\markboth{{\sc Michael Flohr} and {\sc Holger Knuth}}{{\sc On Verlinde-Like Formulas in $c_{p,1}$ LCFTs}}

\address{Institut f\"ur Theoretische Physik, Leibniz Universit\"at Hannover \newline%
\indent Appelstrasse 2, D-30167 Hannover, Germany}%
\email{knuth / flohr @itp.uni-hannover.de}%
\urladdr{http://www.itp.uni-hannover.de /$\sim$knuth resp. /$\sim$flohr}

\section{Introduction}
The connection between the modular transformation properties of the characters in rational conformal field theories (RCFTs) and the fusion algebra of their chiral symmetry algebra is one of the keystones, which have led to our deep mathematical understanding of these compared to other non-trivial quantum field theories. Since its eponymous proposal in 1988 by E. Verlinde \protect\cite{Verlinde:1988sn} the Verlinde-formula provided an enormous simplification to the calculation of the fusion rules. Its proof, however, was first found in the version, in which it is known in algebraic geometry, by Faltings in 1994 \protect\cite{Faltings:1994}. Here it calculates the dimension of the space of holomorphic sections of certain line bundles over a given moduli space. \\
The rigorous connection between these two fields is developed in a research program describing the vertex operator algebras associated with these models by Huang, Lepowsky and others. In this context a proof for the Verlinde formula was finished only recently -- via \protect\cite{huang-2003} -- in \protect\cite{huang-2004} (also cf. \protect\cite{huang-2005-102}).\\
In this paper we will investigate the fusion rules of the $c_{p,1}$ models, which saw the light of day in a series of papers starting with Gurarie's \protect\cite{Gur93} giving the $c_{2,1}=-2$ model as the simplest example for logarithmic conformal field theories. The latter were proposed by Saleur in \protect\cite{Saleur:1991hk} for the description of two-dimensional polymers. Since then much work has been done to develop these models at first especially for $c=-2$ (e.g. \protect\cite{Gaberdiel:1996np,Kausch:1995py,Kau00}), but also for general $p$ (e.g. \protect\cite{Gaberdiel:1996kx,Flo96,Flohr:1996vc}), as reviewed also in \protect\cite{Flo03,Gab03}. Recently some results yet only existing for $c=-2$ could be generalised to all $p\geq 2$, as in \protect\cite{Carqueville:2005nu} or \protect\cite{Flohr:2006id}. Also the study of the vertex operator algebras relevant for these models has advanced remarkably lately (cf. \protect\cite{adamovic-2007}). Feigin et al. have investigated the connections of such $c_{p,1}$ models to quantum groups via a Kazhdan-Lusztig Correspondence (cf. \protect\cite{Feigin:2005xs,Feigin:2005zx}).\\
The $c_{p,1}$ models are also well known to be rational in the weak sense, that their primary fields fall into finitely many blocks with respect to an extended symmetry algebra. They are found to be governed by extensions of the Virasoro algebra at $c=c_{p,1}$ by triplets of fields with integer conformal weight (cf. \protect\cite{Kau91}), the triplet $\mathcal{W}$-algebras $\mathcal{W}(2,(2p-1)^{\otimes 3})$, which  are their maximally extended local chiral symmetry algebras. However, these algebras have reducible but indecomposable representations, which we from now on just call indecomposable representations in contrast to the irreducible ones. So these models are logarithmic conformal field theories (LCFTs) and have a non-semisimple representation category of their vertex algebra. They are not subject to the mentioned proof of the Verlinde formula, for which the conformal field theories has to be rational in the strong sense, where the semisimplicity of the representation category is required. We will refer to this definition of rationality, when we talk about RCFTs. \\
The fusion product representations are defined by the action of the modes of the meromorphic fields on the product of fields given by a comultiplication formula as reviewed in \protect\cite{Gab00} (also cf. \protect\cite{Moore:1988qv,Gaberdiel:1993td,Gaberdiel:1993mt}). For the $c_{p,1}$ models one needs the fusion rules of all possible pairs of indecomposable and irreducible representations of either the Virasoro algebra or the triplet $\mathcal{W}$-algebra. For the Virasoro algebra these are infinitely many fusion products, which have been calculated for the cases of $p=2$, $p=3$ and partially for higher $p$ by Gaberdiel and Kausch in \protect\cite{Gaberdiel:1996kx}. But they all to decompose into finitely many terms. Many fusion products of two irreducible representations decompose into indecomposable representations. Gaberdiel and Kausch's calculation for the triplet algebra for $p=2$ presented in \protect\cite{Gaberdiel:1996np} has proven that the $c_{2,1}=-2$ model is rational in the slightly stronger sense compared to the mentioned weak sense, in which the fusion products are also required to decompose into finitely many direct summands. Furthermore fusion rules for other logarithmic conformal field theories have been calculated in \protect\cite{Eberle:2006zn} and for logarithmic minimal lattice models in \protect\cite{Pearce:2006sz,Pearce:2006we}.\\
More parallels to RCFTs have been found: The partition functions of the $c_{p,1}$ models could be calculated in terms of the characters of irreducible representations and further forms associated to indecomposable ones (cf. \protect\cite{Flo96,Flohr:1996vc}). The latter are not their characters, as these characters are linearly dependent with the ones of the irreducible representations. This is also the major problem, when one tries to calculate a S-matrix to appear in the Verlinde formula. In the RCFT case it gives the transformation of the characters of irreducible representations under one of the generators of the modular group $SL(2,\mathbb{Z})$, $\tau \rightarrow -1/\tau$, which we call $\mathcal{S}$ (and the other generator $\mathcal{T}$). The characters are also referred to as the canonical basis of the vacuum torus amplitudes. For the $c_{p,1}$ models we do not have this canonical basis to calculate the S-matrix, as the indecomposable representation have to be taken care of. But in \protect\cite{Flohr:1996vc} the forms already representing the indecomposable representations in the partition function are used just for this purpose. These forms depend on a parameter, which we call $\alpha$ throughout this paper, and become the characters of indecomposable representations in the limit $\alpha \rightarrow 0$. An adaption of the Verlinde formula is found there, in which this limit has to be taken: the limit-Verlinde formula. It gives the correct fusion rules, as far as they are known from \protect\cite{Gaberdiel:1996kx,Gaberdiel:1996np}, after a manual replacement of combinations of irreducible representation by indecomposable representations, which can not be distinguished by methods based on the modular transformation properties of characters. \\ 
In \protect\cite{Flohr:2005cm} it has been shown for the case $p=2$, that the set of forms used to calculate the S-matrix here also is a basis of the chiral vacuum torus amplitudes. Also strong arguments are presented there in favour of this to be true for all values of $p$. Furthermore the $\text{C}_2$-cofiniteness of the corresponding conformal vertex algebras has been shown first for $p=2$ (cf. \protect\cite{abe-2005}) and then in general (cf. \protect\cite{Carqueville:2005nu}). This is part of the definition of a rational conformal vertex algebra as given in \protect\cite{Fuchs:2006nx}.\\
This latter paper details the argumentation leading to an alternative ''generalised'' Verlinde formula for the $c_{p,1}$ models for fusion products of two irreducible representations, which has been presented by Fuchs et al. in \protect\cite{Fuchs:2003yu}. While in the case of RCFTs the matrices $N_I$ containing the fusion coefficients are diagonalised simultaneously by the S-matrix, they are block diagonalised simultaneously here by the S-matrix of this method, which is found by the construction of an $SL(2,\mathbb{Z})$ representation with the help of an automorphy factor. The result is a fusion algebra, which is closed within the irreducible representations. This fusion algebra again only corresponds to the results of Gaberdiel and Kausch through the same replacements as for the limit-Verlinde formula. 
\\
In section \ref{SecFlohr} the previous work on the limit-Verlinde formula will be reviewed and complemented. Especially the arguments for the needed manual replacements are detailed in section \ref{SecFlohr} and the explicit form of the S-matrix used in this Verlinde formula is given for general $p$. In the appendix we further supplement our discussion. Appendix \ref{appchoice} makes clear that the choice, which one has on the forms representing the indecomposable representations, has no influence on the results of this method whatsoever. Appendix \ref{appfusp2p3} explicitly gives the fusion rules for $p=2$ and $p=3$ before and after the replacements. This is complemented by a demonstration through a few examples, how these replacements come about, in appendix \ref{appDemRRp3}. \\
In section \ref{SecExt} we develop an extension of Fuchs et al.'s work yielding a block diagonalisation method, which also incorporates the indecomposable representations. The correct extension of the S-matrix simultaneously block diagonalising the -- now also larger -- fusion coefficient matrices for both irreducible and indecomposable representations is found in a few steps starting from the mentioned automorphy factor. We also use the known fusion rules for $p=2$ to accomplish this, as it is detailed in appendix \ref{SCKfromp2}. \\
In section \ref{Projection} we proof that this extended block diagonalisation method reduces to its archetype by simply projecting on the components associated to the irreducible representations.\\  
In section \ref{SecEqui} we show that our extended block diagonalisation method gives the same results as the limit-Verlinde formula for all $p \geq 2$, which amounts to equivalence of both approaches. Together with section \ref{Projection} we thus find that all three methods -- the small and extended block diagonalisation method and the limit-Verlinde formula -- compute the same fusion rules for products of two irreducible representations.\\
Finally in section \ref{SecBPZ} we proof explicit expressions for the decompositions of the fusion products for general $p$ in BPZ-like form following from our ''generalised'' Verlinde-formulas. Indeed, the fusion rules have the usual form for CFTs, as it was already seen in the paper \protect\cite{BPZ84} of Belavin and Polyakov and Zamolodchikov, which has laid the headstone of the whole field of conformal field theories. We also apply the replacement rules we have found to these expressions and find for the products of irreducible representations the formula proposed by Gaberdiel and Kausch in \protect\cite{Gaberdiel:1996kx}.\\ 
More details on this work can be found in the diploma thesis of HK \protect\cite{Knuth:2006}.

\section{The Limit-Verlinde Formula} \label{SecFlohr}
The proposal for calculating the fusion rules given as ''case III'' in earlier work of MF (\protect\cite{Flohr:1996vc}) is reviewed and supplemented in this section. As we already mentioned in the introduction, it is based on a S-matrix calculated from a set of forms consisting of the characters of irreducible representations of the triplet $\mathcal{W}$-algebra, $\mathcal{W}(2,(2p-1)^{\otimes 3})$, and further $(p-1)$ forms, which depend on a parameter $\alpha$ and with which the set closes under modular transformations $\gamma \in SL(2,\mathbb{Z})$. In the limit $\alpha \rightarrow 0$ it becomes the linear dependent set of characters of irreducible and indecomposable representations. It is a basis of chiral vacuum torus amplitudes (cf. \protect\cite{Flohr:2005cm}). Its specific choice, which we will use in this section, is detailed in appendix \ref{appchoice}. We also discuss there that the freedom of choice, which we have for the mentioned additional forms, has no influence on the outcome for the fusion rules. We will refer to this set as our chosen basis of vacuum torus amplitudes. Our goal is to calculate the S-matrix, $\mathtt{S}_{p,\alpha}$, which gives the transformation $\tau \rightarrow -1/\tau$ of the elements of this basis, defined by
\begin{equation}\label{Stranschi}
\pmb{\chi}_p(\alpha)\left( -\frac{1}{\tau}\right)
=\mathtt{S}_{p,\alpha}\pmb{\chi}_p(\alpha)(\tau).
\end{equation}
Here the components of the vector $\pmb{\chi}_p(\alpha)$ are the chosen basis elements. $\mathtt{S}_{p,\alpha}$ will then be used in an adapted Verlinde formula to calculate the fusion coefficients.\\
We label the irreducible and indecomposable representations of $\mathcal{W}(2,(2p-1)^{\otimes 3})$ by the conformal weights of their highest weight states given in square brackets. These weights are given by the Kac formula for conformal weights of primary fields in minimal models, which also gives the ones of the $c_{p,1}$ models in an extended Kac table $h_{r,s}$ with $0<r<3$ and $0<s<3p$. With the second line of the Kac table being redundant to the first we label the irreducible representations by $[h_{1,\sigma}]$ with $0< \sigma \leq p$ or $2p\leq \sigma <3p$ and the indecomposable representations $\left[\widetilde{h_{1,\sigma}}\right]$ with $p < \sigma < 2p$, where the tilde marks their reducibility. 
We select a sequence of elements of our basis of vacuum torus amplitudes defining the vector
\begin{eqnarray}\label{charvecalpha}
\pmb{\chi}^t_p(\alpha)&=&( \chi^+_{0,p},
\chi^-_{p,p},\chi^+_{p-1,p},\chi^-_{p-1,p},{\tilde{\chi}}_{p-1,p}(\alpha),\chi^+_{p-2,p},\chi^-_{p-2,p},{\tilde{\chi}}_{p-2,p}(\alpha),\ldots,\\
\nonumber
&&\chi^+_{1,p},\chi^-_{1,p},{\tilde{\chi}}_{1,p}(\alpha)).
\end{eqnarray} 
For the characters, $\chi^+$, $\chi^-$, this corresponds to the sequence chosen in \protect\cite{Fuchs:2003yu}. They are linear combinations of Riemann-Jacobi $\Theta$-functions, $\Theta_{\lambda,p}$, and affine $\Theta$-functions, $(\partial\Theta)_{\lambda,p}$, divided by the Dedekind $\eta$-functions, except for the two projective irreducible modules, which are proportional to a Riemann-Jacobi $\Theta$-functions divided by the Dedekind $\eta$-functions (cf. \protect\cite{Flohr:1996vc}). The additional $(p-1)$ chiral vacuum torus amplitudes, ${\tilde{\chi}}(\alpha)$, contain each one of the forms 
\begin{equation*}
(\nabla \Theta)_{\lambda,p} = \mathrm{i} \tau (\partial \Theta)_{\lambda,k}
= \frac{1}{2 \pi} \log(q)(\partial \Theta)_{\lambda,k}
\end{equation*}
instead of the affine $\Theta$-function. \\
With the vector 
\begin{eqnarray*}
\pmb{\Theta}^t_p &=& \frac{1}{\eta}\biggl(\Theta_{0,p},\Theta_{1,p},\ldots
\Theta_{p,p},(\partial\Theta)_{1,p},(\partial\Theta)_{2,p},\ldots
(\partial\Theta)_{p-1,p},\biggr.\\
&&\biggl. -(\nabla\Theta)_{1,p}, -(\nabla\Theta)_{2,p}, \ldots,
-(\nabla\Theta)_{p-1,p}\biggr)
\end{eqnarray*}
we can express the vector $\pmb{\chi}_p(\alpha)$ by the multiplication of the matrix $B$ of coefficients in the linear combinations and the vector of $\Theta$-functions:
\begin{equation}\label{Bdef}
\pmb{\chi}_p(\alpha) = B \pmb{\Theta}_p
\end{equation}
The matrix $B$ has only few non-zero components:
\begin{eqnarray}\label{Belements}
B_{1,1}=1\, , &\quad &B_{2,p}=1\, , \\
\nonumber B_{3s,p-s}=\frac{s}{p}\, , &\quad&  B_{3s,2p-s+1}=\frac{1}{p}\, , \\
\nonumber B_{3s+1,p-s}=\frac{p-s}{p}\, , &\quad&  B_{3s+1,2p-s+1}=-\frac{1}{p}\, , \\
\nonumber B_{3s+2,p-s}=2\, , &\quad&  B_{3s+2,2p+s}=\mathrm{i} \alpha
\end{eqnarray}
for $0 < s < p$. The last line encodes the choice for the forms ${\tilde{\chi}}(\alpha)$ discussed in appendix \ref{appchoice}. \\
The modular transformation properties of the $\Theta$-functions are known and the transformation $\tau \rightarrow -1/\tau$ of the vector $\pmb{\Theta}_p$ is given by the matrix $\mathfrak{S}$ defined by
\begin{equation*}
\pmb{\Theta}_p\left( -\frac{1}{\tau}\right)
=\mathfrak{S}_{p,\alpha}\pmb{\Theta}_p(\tau)
\end{equation*}
This matrix has three non-zero blocks with the components
\begin{eqnarray}
\label{Thetatrans} \begin{aligned}
\mathfrak{S}_{ij}&=
\frac{1}{1+\delta_{j,1}+\delta_{j,p+1}} \sqrt{\frac{2}{p}}
\cos\left(\frac{\pi (i-1) (j-1)}{p}\right)&
\quad &\forall \, 0<i,j \leq p+1\; ,\\ 
\mathfrak{S}_{(2p+k)(p+l+1)}&= \mathrm{i} \sqrt{\frac{2}{p}} 
\sin\left(\frac{\pi k l}{p}\right)&
\quad &\forall \, 0<k,l<p\; ,\\
\nonumber \mathfrak{S}_{(p+n+1)(2p+m)}&= -\mathrm{i} \sqrt{\frac{2}{p}}
\sin\left(\frac{\pi n m}{p}\right)& \quad &\forall \, 0<n,m<p\; .
\end{aligned}
\end{eqnarray}
So finally together with equations \eqref{Bdef} and \eqref{Stranschi} the S-matrix, $\mathtt{S}_{p,\alpha}$, is equal to the matrix product $B\mathfrak{S}B^{-1}$. This leads to a block structure with one $2 \times 2$ block,
$S(p)_{0,0}$, and each $(p-1)$ $2 \times 3$ and $3 \times 2$ blocks,
$S(p)_{0,l}$ and $S(p)_{s,0}$, respectively. These blocks do not
depend on $\alpha$. The rest of the matrix is filled with $3
\times 3$ blocks $S(p,\alpha)_{s,l}$. The indices $s$ and $l$ always take values between $1$ and $p-1$, inclusively, and $p \geq 2$. We get:
\begin{equation}\label{FlohrS}
\mathtt{S}_{p,\alpha}=\begin{pmatrix}
S(p)_{0,0} & S(p)_{0,1} & \dotso & S(p)_{0,p-1}  \\
S(p)_{1,0} & S(p,\alpha)_{1,1} & \dotso & S(p,\alpha)_{1,p-1} \\
\vdots & \vdots & \ddots & \vdots &\\
S(p)_{p-1,0} & S(p,\alpha)_{p-1,1} & \dotso & S(p,\alpha)_{p-1,p-1}
\end{pmatrix}
 \end{equation}
 with
\begin{eqnarray*}
 S(p)_{0,0}&=&\frac{1}{\sqrt{2p}}
\begin{pmatrix}
1 & 1 \\
1 & (-1)^p
                           \end{pmatrix}\; ,\\
 S(p)_{0,l}&=&\frac{2}{\sqrt{2p}}\begin{pmatrix}
1 & 1 &0\\
(-1)^{p-l} & (-1)^{p-l} &0
                           \end{pmatrix}\; ,\\
 S(p)_{s,0}&=&\frac{1}{\sqrt{2p}}\begin{pmatrix}
 \frac{s}{p} & (-1)^{p+s}\frac{s}{p} \\
 \frac{p-s}{p} & (-1)^{p+s}\frac{p-s}{p}\\
 2 & 2(-1)^{p+s}
\end{pmatrix}\; ,
\end{eqnarray*}
\begin{eqnarray*}
S(p,\alpha)_{s,l}&=&\frac{2}{\sqrt{2p}}(-1)^{p+l+s}\times \\
 &&\begin{pmatrix}
 \frac{s}{p} \mathfrak{c}_{sl} + \frac{2}{p}
\frac{1}{\alpha} \mathfrak{s}_{sl}& \frac{s}{p} \mathfrak{c}_{sl}
+ \frac{2}{p} \frac{1}{\alpha} \mathfrak{s}_{sl} & -\frac{1}{p
\alpha}
\mathfrak{s}_{sl} \\
 \frac{p-s}{p} \mathfrak{c}_{sl} - \frac{2}{p}
\frac{1}{\alpha} \mathfrak{s}_{sl} & \frac{p-s}{p}
\mathfrak{c}_{sl} - \frac{2}{p} \frac{1}{\alpha} \mathfrak{s}_{sl}
& \frac{1}{p \alpha}
\mathfrak{s}_{sl} \\
 2 \mathfrak{c}_{sl} - \alpha (p-l)
\mathfrak{s}_{sl} &  2 \mathfrak{c}_{sl} + \alpha l
\mathfrak{s}_{sl} & 0 \end{pmatrix}
\end{eqnarray*}
with the abbreviations $\mathfrak{c}_{sl}=\cos{\left(\pi
\frac{sl}{p}\right)}$ and $\mathfrak{s}_{sl}=\sin{\left(\pi
\frac{sl}{p}\right)}$.\\
This matrix fulfils ${\mathtt{S}_{p,\alpha}}^2=\mathtt{1\!\!l}$,
but is not symmetric. For $\alpha \rightarrow 0$ the forms
${\tilde{\chi}}_{\lambda,p}(\alpha)$ pass into the characters of
the indecomposable representations. So they are linearly dependent
with the characters of the irreducible representations in this
limit. Consequently some of the entries of
$\mathtt{S}_{p,\alpha}$ diverge in this case.\\
For completeness the matrix $\mathtt{T}_{p,\alpha}$ for the
transformation $\tau\rightarrow \tau+1$ is given here. It is
defined as
\begin{equation}
\label{Ttranschi} \pmb{\chi}_p(\alpha)(\tau+1)
=\mathtt{T}_{p,\alpha}\pmb{\chi}_p(\alpha)(\tau)
\end{equation}
and is calculated to be
\begin{eqnarray*}
{\mathtt{T}_{p,\alpha}}(\tau)&=& T(p)_{0,0} \oplus
\bigoplus^{p-1}_{s=1} T(p,\alpha)_{s,s}\; ,\\
T(p)_{0,0}&=&
\begin{pmatrix}
 \mathrm{e}^{-\mathrm{i}\frac{\pi}{12}} & 0 \\
 0 & \mathrm{e}^{-\mathrm{i}\pi\left(\frac{p}{2}-\frac{1}{12}\right)}
                           \end{pmatrix}\; ,\\
T(p,\alpha)_{s,s}&=&
\begin{pmatrix}
\mathfrak{t}_s& 0 & 0 \\
0 & \mathfrak{t}_s & 0 \\
\mathrm{i} \alpha \,(p-s) \,\mathfrak{t}_s &  -\mathrm{i} \alpha \,s
\,\mathfrak{t}_s & \mathfrak{t}_s \end{pmatrix} \end{eqnarray*}
with
\begin{equation*}
\mathfrak{t}_s=\mathrm{e}^{-\mathrm{i}\pi\left(\frac{(p-s)^2}{2p}-\frac{1}{12}\right)}\; .
\end{equation*}
The matrices $\mathtt{S}_{p,\alpha}$ and
$\mathtt{T}_{p,\alpha}$ describe the action of the generators
$\mathcal{S}$ and $\mathcal{T}$ of the modular group
$SL(2,\mathbb{Z})$ on $\pmb{\chi}_p(\alpha)(\tau)$. So with
equations \eqref{Stranschi} and \eqref{Ttranschi} any element
$\gamma \in SL(2,\mathbb{Z})$ can be represented as a matrix
$\mathtt{G}_{p,\alpha}(\gamma)$, which is a product only
containing copies of  $\mathtt{S}_{p,\alpha}$ and
$\mathtt{T}_{p,\alpha}$, such that
\begin{equation}
\label{GalphaRep} \pmb{\chi}_p(\alpha)\left( \gamma\tau\right)
=\mathtt{G}_{p,\alpha}(\gamma)\pmb{\chi}_p(\alpha)(\tau)
\end{equation}
As the action of $SL(2,\mathbb{Z})$ on functions on $\mathbb{C}$
is linear\footnote{The action of $SL(2,\mathbb{Z})$ on a function
$f: \mathbb{C}\rightarrow\mathbb{C}$ shall be defined as the
composition $f\circ\gamma$ with $\gamma \in SL(2,\mathbb{Z})$.},
we directly have for two elements $\gamma,\gamma'  \in
SL(2,\mathbb{Z})$, that
\begin{equation}
\label{GalphaHom} \mathtt{G}_{p,\alpha}(\gamma\gamma') =
\mathtt{G}_{p,\alpha}(\gamma) \mathtt{G}_{p,\alpha}(\gamma')
\end{equation}
It follows that $\mathtt{S}_{p,\alpha}$ and
$\mathtt{T}_{p,\alpha}$ generate a representation of
$SL(2,\mathbb{Z})$, namely $\mathtt{G}_{p,\alpha}(\gamma)$, for
a fixed $\alpha \neq 0$. We can also immediately see that like
the generators of $SL(2,\mathbb{Z})$ also
$\mathtt{S}_{p,\alpha}$ and $\mathtt{T}_{p,\alpha}$ have to
fulfil the conditions
${\mathtt{S}_{p,\alpha}}^2=\mathtt{1\!\!l}$ and
${(\mathtt{S}_{p,\alpha}\mathtt{T}_{p,\alpha})}^3=\mathtt{1\!\!l}$. As
an easy check one can calculate these products for any $p$, which
we did for up to $p=6$. \\
The matrix $\mathtt{S}_{p,\alpha}$ is now plugged into the
Verlinde formula as known for rational conformal field theories.
This, of course, leads to an object ${\mathtt{N}_{ij}}^k(\alpha)$,
which depends on $\alpha$. But here the limit of $\alpha
\rightarrow 0$ exists, as we will proof for all $p$ along with our outcome of the comparison of this approach with the block diagonalisation method in section \ref{SecEqui}. We define the coefficients
${\mathtt{N}_{ij}}^k$ to be exactly this limit and get the
limit-Verlinde formula:
\begin{equation}
\label{Nijkalpha} {\mathtt{N}_{ij}}^k=\lim_{\alpha\rightarrow
0}{\mathtt{N}_{ij}}^k(\alpha) =\lim_{\alpha\rightarrow 0}\biggl(
\sum_{r=1}^{3p}\frac{(\mathtt{S}_{p,\alpha})_{jr}(\mathtt{S}_{p,\alpha})_{ir}{(\mathtt{S}_{p,\alpha})_r}^k}{(\mathtt{S}_{p,\alpha})_{3,r}}\biggr)
\end{equation}
Note that the third component of the vector
$\pmb{\chi}_p(\alpha)(\tau)$ is the character of the vacuum
representation. In contrast to the semisimple case of RCFTs with
symmetric S-matrix, the indices of $\mathtt{S}_{p,\alpha}$ in
the Verlinde formula have to be kept as in this formula.
Especially the third line of $\mathtt{S}_{p,\alpha}$ -- rather
than the column -- has to be taken for the denominator of the
$\alpha$-Verlinde formula. This is due to a convention of
left-multiplication of $\mathtt{S}_{p,\alpha}$ with
$\pmb{\chi}_p(\alpha)(\tau)$ in eq. \eqref{Stranschi}, which we have
chosen quite naturally.\\
At first sight the results for ${\mathtt{N}_{ij}}^k$ for $p=2$ and
$p=3$, which are given in the appendix in tables \ref{prefusp2} and
\ref{prefusp3}, differ quite much from the fusion coefficients
calculated in \protect\cite{Gaberdiel:1996np} and
\protect\cite{Gaberdiel:1996kx}. However, we have to note that any fusion
rules we get using eq. \eqref{Nijkalpha} by itself, can only be
taken as true on the level of characters, not representations,
because the calculation is based only on the modular
transformation properties of the characters. Here we have the
problem, that, as soon as we take the limit $\alpha \rightarrow
0$, the functions ${\tilde{\chi}}_{\lambda,p}(\alpha)$ become the characters of the indecomposable representations, which are a linear combination of characters of irreducible
representations. To be precise the relation between the characters of the irreducible representations, $\chi^+_{\lambda,p}$ and
$\chi^-_{\lambda,p}$, and those of the indecomposable representations, $\chi^{\mathcal{R}}_{\lambda,p}$, is
\begin{equation}
\label{charlincomb}
 2 \chi^+_{\lambda,p} + 2 \chi^-_{\lambda,p} = \frac{2}{\eta(\tau)}
    \Theta_{\lambda,p} =
 \chi^{\mathcal{R}}_{\lambda,p}
\end{equation}
for $0<\lambda<p$.\\
So the method
presented here can not distinguish the indecomposable
representation from these linear combination of irreducible
representations in the decomposition of
the fusion product in the first place.\\
Indeed, for many fusion products there are 
components ${\mathtt{N}_{ij}}^k$ corresponding to these linear combinations, while
in \protect\cite{Gaberdiel:1996np} and \protect\cite{Gaberdiel:1996kx} the
corresponding indecomposable representation have been found to be the correct
result. \\
There is another problem that occurs in fusion products of
indecomposable representations with some other representation: For
one and the same fusion product ${\mathtt{N}_{ij}}^k$ encodes both the
linear combinations mentioned above and the
corresponding indecomposable representations, which then have a
negative integer coefficients. These problems are illustrated
in the case of $p=3$ in appendix \ref{appDemRRp3}.\\
Without clear rules for these replacements the value of the results would be lost. 
Fortunately the triplet algebra, $\mathcal{W}(2,(2p-1)^{\otimes 3}$, has a rescaled $\mathfrak{su}(2)$ subalgebra, which is formed by the zero modes of the fields $W^{(a)}$ extending the Virasoro algebra to $\mathcal{W}(2,(2p-1)^{\otimes 3}$. We know that fusion products of irreducible representations can only decompose into irreducible
representations, which have the correct $\mathfrak{su}(2)$ quantum number $j$ with respect to this subalgebra and any indecomposable representations, because they have no
unique $\mathfrak{su}(2)$ quantum numbers. \\
This rules out all the combinations of both singlets and doublets,
with respect to $j$, in the decomposition. The two irreducible representations having
the characters on the left hand side of equation
\eqref{charlincomb}, which gives our ''translation'' to the
correct fusion rules, are exactly a singlet and a doublet and thus
forbidden. This justifies the permanent replacement in all fusion
products of two irreducible representations, in which the mentioned
combinations appear. \\
The products with indecomposable representations are a bigger
problem because the argument of $\mathfrak{su}(2)$ quantum numbers can not be
applied, when the representation does not have unique quantum
numbers. Here a practical argument is given by the negative
coefficients. These should be mended, which seems to be possible
for all $p$ as well. \\
There are quite a few fusion products still left out, in
which a replacement should be made, but where we have no argument
except the result. For example, for $p=3$ there are 7 fusion
products of this kind left (cf. table \ref{prefusp3} in app. \ref{appfusp2p3}). But
there is also no argument, why exactly these
should be exceptions.\\
All in all we can surely say that the following rules are
well-founded. There are no indications of deviations whatsoever:
\begin{itemize}
\item Replace the left hand side of the following ''equation'' by the indecomposable representation on the right hand side, whenever it
appears:
\begin{equation}
\label{replincomb} 2\left[h_{1,p-\lambda}\right]+2\left[h_{1,3p-\lambda}\right]=\left[\widetilde{h_{1,p+\lambda}}\right]
\quad \lambda=1\ldots p-1\quad .
\end{equation}
\item If two coefficients appear now for the same indecomposable representation in one fusion rule, add
them.
\end{itemize}
If there is a negative coefficient of an indecomposable
representation in the decomposition of the fusion product, it has
to be compensated by a higher positive multiplicity from the first
rule to make sense. We checked this up to $p=6$. \\
Finally the following conjecture summarises this method.
\liketheorem{Conjecture:}{} The structure constants,
${\mathcal{N}_{ij}}^k$, of the fusion algebra of the $c_{p,1}$
series are calculated by equation \eqref{Nijkalpha} for all
$i=1\ldots(3p-1)$ and
\begin{itemize}
\item for all $(j,k) \in \lbrace 1,2\rbrace\times\lbrace 1,\ldots,3p-1\rbrace$ and all $(j,k) \in \lbrace 3,\ldots,3p-1\rbrace\times\lbrace 1,2\rbrace$ as
${\mathcal{N}_{ij}}^k={\mathtt{N}_{ij}}^k$
\item for all $(j,k) \in \lbrace 3,3p-1\rbrace\times\lbrace k \in \lbrace 3,\ldots,3p-1\rbrace \vert k \mod 3= 0 \rbrace$ and $\kappa=k,(k+1)$ as
\begin{eqnarray*}
{\mathcal{N}_{ij}}^\kappa&=& \left\{ \begin{aligned}
0 \quad & \text{if} \quad {\mathtt{N}_{ij}}^k={\mathtt{N}_{ij}}^{(k+1)} \\
{\mathtt{N}_{ij}}^{\kappa} \quad & \text{else}
                           \end{aligned} \right.\quad \quad ,\\
{\mathcal{N}_{ij}}^{(k+2)}&=& \left\{ \begin{aligned}
{\mathtt{N}_{ij}}^{(k+2)}+{\mathtt{N}_{ij}}^k/2 \quad & \text{if} \; \; \; {\mathtt{N}_{ij}}^k={\mathtt{N}_{ij}}^{(k+1)} \\
{\mathtt{N}_{ij}}^{(k+2)} \quad \quad & \text{else}
                           \end{aligned} \right.\quad \quad .
\end{eqnarray*}
\end{itemize} \qef
Here we have stated the proposed connection between the fusion
coefficients ${\mathcal{N}_{ij}}^k$ and the \emph{pre-fusion}
coefficients ${\mathtt{N}_{ij}}^k$, which enables us to compute
the former for any $p$ with little expenses. However, the limit in
this procedure makes it hard to understand the cause, why this
leads to the correct result. The situation looks surely a bit
better after the work in \protect\cite{Flohr:2005cm} gave us the new
perspective on the functions ${\tilde{\chi}}_{\lambda,p}(\alpha)$
as chiral vacuum torus amplitudes. But still one advantage of a
different method, which we will discuss in the next section, is
the absence of such a limit. \\
As mentioned above the ambiguities about the indecomposable
representations are generic for methods based on modular
transformation properties of characters. So there is virtually no
hope to find a method using some kind of Verlinde formula, which
does not exhibit them. But this is something we gladly cope with,
as the limit-Verlinde formula reduces the amount of needed
calculation to get the fusion rules for any particular $p$
enormously.

\section{Block Diagonalisation of the Fusion Rules}
\label{SecExt} In this section we present an extension of the approach of Fuchs et al., first published in \protect\cite{Fuchs:2003yu}, now  including the indecomposable representations as well. First we want to mention a few key
features already in the beginning. A limit like in the last section does not appear. This
method is motivated by the statement that any non-semisimple,
finitely generated, associative and commutative algebra, like the
fusion algebra we look for here, is the direct sum of its radical
and some semisimple algebra. As the key consequence a matrix
$P_{p}$ is found, which simultaneously block diagonalises the
matrices $N_{p,I}$ of pre-fusion coefficients in contrast to the case of RCFTs,
where the fusion algebra of the Virasoro irreducible modules is
semisimple and the S-matrix diagonalises the fusion
coefficient matrices $N_I$ simultaneously. We first find out, how the simultaneous block diagonalisation
comes about and see that the matrix $P_{p}$ is a matrix
consisting of simultaneous eigenvectors of the matrices
$N_{p,I}$. Afterwards we find a S-matrix $S_p$ and the extension of all other matrices appearing in the original block diagonalisation approach in \protect\cite{Fuchs:2003yu}. While large parts of the argumentation in that paper were in a general
setting for non-semisimple fusion algebras, we will restrict
ourselves here to the case of the $c_{p,1}$ models using the same notation.

\subsection{Simultaneous Eigen Decomposition of the Fusion Coefficient Matrices}
This subsection follows the argumentation of its archetype by Fuchs et al. (\protect\cite{Fuchs:2003yu}) quite closely.\\ 
We want to block diagonalise the matrices of
\mbox{pre-fusion} coefficients for the full \mbox{pre-fusion}
algebra including indecomposable representations simultaneously.
This pre-fusion algebra is defined in the familiar way:
\begin{equation}
\label{fusExtalg} X_I X_J = \sum_{K=1}^{3p-1} {(N_{p})_{IJ}}^K X_K
\; .
\end{equation}
The basis $X$ is now larger than in \protect\cite{Fuchs:2003yu} and also contains the indecomposable
representations. Its sequence is chosen to be the same as the one of the vector $\pmb{\chi}_p(\alpha)$ (eq. \eqref{charvecalpha}). \\
We now change the basis  in view of
the direct sum of a semisimple algebra and a radical, which is
equal to the pre-fusion algebra. The new one consists of the union of
a set of primitive idempotents, $e_A$ with $A=1\ldots p+1$, in the
semisimple algebra and a basis of the radical, $w_A$ and $w'_A$ with $A=3\ldots p+1$. All the primitive idempotents $e_A$ form a
partition of the unit element of the semisimple algebra (and also
the whole pre-fusion algebra):
\begin{equation}
\label{part1} \sum_{A=1}^{p+1} e_A = \mathtt{1\!\!l}\; ,
\end{equation}
Each pair $w_A$ and $w'_A$ corresponds to an idempotent, $e_A$, with an image of dimension 3.
There are two further primitive idempotents in the new basis with
a one dimensional image ($A=1,2$). The new basis, called $Y$, is
taken in the following order:
\begin{equation*}
Y=(e_1,e_2,e_3,w_3,w'_3,e_4,w_4,w'_4,\ldots,e_{p+1},w_{p+1},w'_{p+1})\;  .
\end{equation*}
The idempotents of the semisimple algebra and the basis of the radical relate to each other by
\begin{eqnarray}
\label{relYa}
e_A e_B &=& \delta_{A,B} e_B\; , \\
e_A w_C &=& \delta_{A,C} w_C\; , \\
e_A w'_C &=& \delta_{A,C} w'_C\; ,\\
w_C w_D &=& 0\; ,\\
w'_C  w_D &=& 0\; ,\\
\label{relYe} w'_C w'_D &=& 0
\end{eqnarray}
with $0 < A,B \leq p+1$, $3 \leq C,D \leq p+1$ and $\delta$ being the Kronecker delta.\\
The change of basis is given by $P_{p}$ defined by
\begin{equation}
\label{Pdef}
 X_L = \sum_{J=1}^{3p-1} {(P_{p})_L}^J Y_J \; .
\end{equation}
We will see in the following proposition and its proof that this matrix is the essential entity to be calculated, as the pre-fusion coefficients can be expressed in terms of its matrix elements only.
\liketheorem{Proposition:}{} $P_{p}$
block diagonalises the matrices $N_{p,I}$ simultaneously, i.e.
\begin{equation}
\label{NPMP}
 N_{p,I}=P_{p} M_{p,I} {P_{p}}^{-1}
\end{equation}
with block diagonal matrices $M_{p,I}$, $0< I \leq 3p-1$. The $I$-th row of $P_{p}$, $\pi_I$, is
related to the row corresponding to the vacuum representation,
$\pi_{\Omega}$, by
\begin{equation}
\label{rowPMExt}
 \pi_I = \pi_{\Omega} M_{p,I}
\end{equation}
for all $0< I \leq 3p-1$. \qef \likeremark{Remark:}{} We will proof
these statements, as we calculate now an explicit expression for
$M_{p,I}$ in terms of matrix elements of $P_{p}$.
\proof{$\!\!\!\!\!\!$:} 
On the one hand we multiply equation \eqref{fusExtalg}
by ${({P_{p}}^{-1})_L}^J$ and sum over $J$:
\begin{eqnarray}
 \nonumber X_I Y_L &=& \sum_{K,J,R,S = 1}^{3p-1} {({P_{p}}^{-1})_L}^J {(N_{p})_{IJ}}^K {(P_{p})_K}^S {({P_{p}}^{-1})_S}^R X_R \\ \label{implMI}
 &=& \sum_{K,J,S = 1}^{3p-1} \underbrace{{({P_{p}}^{-1})_L}^J {(N_{p})_{IJ}}^K {(P_{p})_K}^S}_{=:{(M_{p,I})_L}^S}
Y_S\; .
\end{eqnarray}
Hence the matrices $M_{p,I}$ give the decompositions of the
products of $X_I$ and $Y_L$ into linear combinations of $Y_S$ for
$I,L=1\ldots 3p-1$.\\
On the other hand with the relations between the elements of the basis $Y$ (eqs. \eqref{relYa}-\eqref{relYe}) and equation
\eqref{Pdef} one can calculate the product on the left hand side
\begin{equation}
\label{XYProd}
  X_I Y_A =\left\{ \begin{aligned}
{(P_{p})_{IA}} Y_A \quad &\text{for}& A=&1,2\\
{(P_{p})_{IA}} Y_A + {(P_{p})_{I(A+1)}} Y_{A+1} + {(P_{p})_{I(A+2)}} Y_{A+2} \quad &\text{for}& A=&3,6,9,\ldots\\
{(P_{p})_{I(A-1)}} Y_A \quad &\text{for}& A=&4,7,10,\ldots\\
{(P_{p})_{I(A-2)}} Y_A \quad &\text{for}& A=&5,8,11,\ldots \quad
\quad .
                           \end{aligned} \right.
\end{equation}
So $M_{p,I}$ is an upper-triangular block diagonal matrix with all but one $3 \times 3$ blocks and reads
 \begin{eqnarray}
 \label{defMp}
   M_{p,I}&=& M_{p,I,0} \oplus \bigoplus_{n=1}^{p-1} M_{p,I,n}\; ,\\
\nonumber   M_{p,I,0}&=&\begin{pmatrix}
     (P_{p})_{I1} & 0 \\
     0 & {(P_{p})_{I2}}
     \end{pmatrix}\; ,\\
\nonumber   M_{p,I,n}&=&\begin{pmatrix}
     {(P_{p})_{I(3n)}} & {(P_{p})_{I(3n+1)}} & {(P_{p})_{I(3n+2)}}\\
     0 & {(P_{p})_{I(3n)}} & 0\\
    0 & 0 & {(P_{p})_{I(3n)}}
   \end{pmatrix}\; .
 \end{eqnarray}
Now we still need to show the second half of our proposition. The row $\pi_{\Omega}$ of the matrix $P_{p}$ is determined by the fact that the vacuum representation is the unit element of the fusion algebra. Thus eq. \eqref{part1} tells us that the sum of
all idempotents $e_A$ is just the vacuum representation. Eq.
\eqref{Pdef} for the case of the vacuum, $L=\Omega$, reads
\begin{equation*}
X_{\Omega} = \sum_{K=1}^{3p-1} (\pi_{\Omega})^K Y_K \; .
\end{equation*}
A comparison to eq. \eqref{part1}, with the order of the basis $Y$ kept in mind, yields
\begin{equation}
\label{Pvacrow} \pi_{\Omega}=(1,1,1,0,0,1,0,0,\ldots,1,0,0) \; .
\end{equation}
One can plug this into eq. \eqref{fusExtalg} with $X_J$ being the vacuum representation:
\begin{equation}
\label{FusVac} X_I = X_{\Omega} X_I = \sum_{K=1}^{3p-1}
(\pi_{\Omega})^K Y_K X_I \; .
\end{equation}
Because of the commutativity of the algebra we can plug eq.
\eqref{implMI} into eq. \eqref{FusVac}:
\begin{equation}
\label{rowrel}
 X_I = \sum_{K,L=1}^{3p-1} \underbrace{(\pi_{\Omega})^K {(M_{p,I})_K}^L}_{=(\pi_I)^L}
Y_L\; .
\end{equation}
Comparing with the definition of $P_{p}$  (eq. \eqref{Pdef})
equation \eqref{rowPMExt} has been shown.\qed

What is actually done here is a
simultaneous eigen decomposition for the set of matrices $N_{p,I}$. This is possible, because they happen to be the structure constants of the algebra in equation \eqref{fusExtalg} and so are related to each other by the properties of this algebra like commutativity.
This enters our proof, as we plug in \eqref{fusExtalg} at
one point and interchange elements of X. \\
The eigen decomposition is nicely encoded in eqns.
\eqref{relYa}-\eqref{relYe}. Using these relations we can write the structure constants ${(N_{p,I})_j}^k$ in the form of eq. \eqref{NPMP} and calculate for a column $p_J$ of $P_p$ that
\begin{eqnarray}
\label{eigvectproofext}
 N_{p,I} p_J &=& P_p M_{p,I} {P_p}^{-1} p_J = P_p M_{p,I} e_J \\
      \nonumber       &=& \left\{ \begin{aligned}
\left({(P_{p})_{IJ}} p_{J-1} + {(P_{p})_{I(J-1)}} p_J \right)  \quad &\text{for}& J=&4,7,10,\ldots \\
\left({(P_{p})_{IJ}} p_{J-2} + {(P_{p})_{I(J-2)}} p_J \right)  \quad &\text{for}& J=&5,8,11,\ldots \\
{(P_{p})_{IJ}} p_J \quad &\text{else}
                           \end{aligned} \right.\; ,
\end{eqnarray}
where $e_J$ is the $J$-th element of the canonical basis. Using this result one also finds
\begin{eqnarray*}
  (N_{p,I} - (P_p)_{I,J-1} \mathtt{1\!\!l})^2 p_J &=& 0 \quad \text{ for } J=4,7,10,\ldots  \quad ,\\
   (N_{p,I} - (P_p)_{I,J-2} \mathtt{1\!\!l})^2 p_J&=&0 \quad \text{ for }
   J=5,8,11,\ldots \quad .
\end{eqnarray*}
Thus $p_J$, $p_{J+1}$ and $p_{J+2}$ form a simultaneous three dimensional eigenspace with eigenvalues ${(P_p)}_{IJ}$ of the respective matrices $N_{p,I}$ for all $0<I<3p-1$ and $J=3,6,9\ldots$. \\ It also shows that the matrix $P_p$ is a matrix consisting of simultaneous generalised eigenvectors of the matrices  $N_{p,I}$ for all $0<I<3p$.

\subsection{The Connection between $\mathtt{S}_{p,\alpha}$ and $P_p$}\label{SecConSP}
We now need to calculate the matrix $P_p$ of simultaneous eigenvectors of the pre-fusion coefficient matrices $N_{p,I}$. The bottom of the line is that it is connected to the matrix $\mathtt{S}_{p,\alpha}$ from section \ref{SecFlohr} in much the way the $\tau$-dependent S-matrix $\boldsymbol{S}_p(\tau)$ defined in \protect\cite{Fuchs:2003yu} is connected to the corresponding matrix of simultaneous eigenvectors there. $\boldsymbol{S}_p(\tau)$ is given by the transformation of the characters of irreducible representations, $\pmb{\chi}_{irr,p}(\tau)$ (defined by omitting the additional forms in eq. \ref{charvecalpha}), under $\tau \rightarrow -1/\tau$ (explicitly given by eq. (3.4) in \protect\cite{Fuchs:2003yu}). To be more precise the modular transformations of these characters are expressed in the form of $\tau$-dependent matrices,
\begin{equation}
\label{DefTauS}
\pmb{\chi}_{irr,p}\left(\gamma \tau\right)=\mathbf{G}_p(\gamma,\tau)\pmb{\chi}_{irr,p}(\tau)\; ,
\end{equation}
and $\boldsymbol{S}_p(\tau)=\mathbf{G}_p(\mathcal{S},\tau)$. 
The construction of an $SL(2,\mathbb{Z})$ representation with the help of an automorphy factor $j_{p}(\gamma,\tau)$ with $\gamma \in SL(2,\mathbb{Z})$,
\begin{equation}
\rho(\gamma)=j_{p}(\gamma,\tau)\boldsymbol{G}_p(\gamma,\tau)\; ,
\end{equation}
leads to a $\tau$-independent S-matrix (cf. section 4.3 of \protect\cite{Fuchs:2003yu}):
\begin{equation}
\label{SirrDef}
\mathsf{S}(p)=j_{p}(\mathcal{S},\tau)\boldsymbol{S}_p(\tau)
\end{equation}
Here we replace the automorphy factor by a conjugation with the matrix $C_{irr,p}(\tau)$, so that $\mathsf{S}(p) = C_{irr,p}(-1/\tau) \boldsymbol{S}_p(\tau) {C_{irr,p}}^{-1}(\tau)$ and see that it corresponds to a matrix $C_p(\alpha)$ in the same way as $\boldsymbol{S}_p(\tau)$ corresponds to $\mathtt{S}_{p,\alpha}$. But this only partially determines the matrices $C_p(\alpha)$. Through a longer study of the case $p=2$, which is described in appendix \ref{SCKfromp2}, we find the missing matrix entries and also get our $\alpha$-independent S-matrices
\begin{equation}\label{SCalphaSCinvalpha}
S_p=C_p(\alpha)\mathtt{S}_{p,\alpha}{C_p}^{-1}(\alpha)\; ,
\end{equation}
which are directly related to the matrices $P_p$ analogous to the situation in \protect\cite{Fuchs:2003yu}:
\begin{equation}
\label{PSK}
 P_p = S_p K_p\; .
\end{equation}
The matrices $K_p$ turn out to be a simple extension of the corresponding matrices in \protect\cite{Fuchs:2003yu}, which we can extend with ones on the diagonal and zeros for all the additional off-diagonal matrix elements. \\
This can be seen following the argumentation of \protect\cite{Fuchs:2003yu} once again. 
We expect $K_p$ to have the corresponding block diagonal
structure because $S_p$ should block diagonalise the pre-fusion
coefficient matrices. \\
As in \protect\cite{Fuchs:2003yu} conditions on $K_p$ result from the relation of the two known vacuum rows of $S(p)$ (eq. \eqref{SExtBlock}) and
$P_{p}$ (eq. \eqref{Pvacrow}), which $K_{p}$ has to connect. Now the matrix elements, $(S(p)_{1,j})_{1,3}$, of the vacuum row are zero. Thus any element of the third row of a block of $K_p$ is multiplied by zero and does not contribute to the vacuum row of $P_{p}$. However, this way the other elements are restricted in the same way as in \protect\cite{Fuchs:2003yu}. \\
We set the additional third column in the first two rows of each block to zero. This gives us the correct result for the vacuum row of $P_p$ and also is compatible with our goal to be able to reduce the whole extended
method back to its archetype for products of irreducible representations by projection on the
$2p$ components of our basis, which represent the irreducible
representations, as we see in the next section. This provides us also with a reason to use the same normalisation for the four matrix elements of each $3 \times 3$ block, which this projection leaves behind, and take the determinant of this $2 \times 2$ block equal to one. But we also ask the $3 \times 3$ blocks to have determinant one, which fixes the third diagonal element of each block to be one. We are left with 
\begin{eqnarray}
\label{KExtDef}
 K_{p} &=&
   (K_{p})_0 \oplus \bigoplus_{s=1}^{p-1} (K_{p})_s\\
\nonumber (K_{p})_0 &:=& \begin{pmatrix}
                   \frac{1}{{(S_p)_{\Omega}}^1} & 0 \\
           0 & \frac{1}{{(S_p)_{\Omega}}^2}
                \end{pmatrix}\\
\nonumber (K_{irr,p})_s &:=&  \begin{pmatrix}
 \frac{\; 1}{{(S_p)_{\Omega}}^{2s+1} - {(S_p)_{\Omega}}^{2s+2}} &  -{(S_p)_{\Omega}}^{2s+2} & 0\\
 \frac{-1}{{(S_p)_{\Omega}}^{2s+1} -
{(S_p)_{\Omega}}^{2s+2}} &
\frac{1}{{(S_p)_{\Omega}}^{2s+1}} &0 \\
{k^{(s)}}_{1} & {k^{(s)}}_{2} & 1
                \end{pmatrix}\;
\end{eqnarray}
Two matrix elements per block, ${k^{(s)}}_{1} $ and ${k^{(s)}}_{2} $, are left open, which we set to zero, so that the third row of each block is $(0,0,1)$.

\subsubsection{A Replacement for the Automorphy Factor for $\gamma=\mathcal{S}$}
\label{ReplAutom} Concerning the conformal field theory we only know that the matrix
$\mathsf{S}(p)$ (eq. \eqref{SirrDef}) is the one corresponding to the
$\mathcal{S}$-transformation, $\tau \rightarrow - \frac{1}{\tau}$, which results from the construction of a representation of the modular group, $SL(2,\mathbb{Z})$, from the modular transformation properties of the characters of the irreducible
representations. To accomplish this an automorphy factor is
needed. But an additional interpretation giving a more direct
connection to physically relevant quantities or properties would
be favourable. This has been the motivation to find a matrix
$C_{irr,p}(\tau)$, which almost conjugates\footnote{Conjugation is
always meant in a group theoretical sense -- not complex conjugate
or suchlike. We say, a matrix $M$ conjugates two (similar)
matrices $N_1$ and $N_2$, if $N_1=M N_2 M^{-1}$.} -- we need a
small alteration due to the $\tau$ dependence of $C_{irr,p}(\tau)$
-- the two matrices $\boldsymbol{S}_p(\tau)$ and $\mathsf{S}(p)$
and replaces the automorphy factor. In this way we see
$\mathsf{S}(p)$ as the matrix giving the
$\mathcal{S}$-transformation of \mbox{$\tau$-dependent} linear
combinations, $\mathbf{\chi'}_{irr,p}(\tau)$, of characters,
$\mathbf{\chi}_{irr,p}(\tau)$, of irreducible representations given by $C_{irr,p}(\tau)$:
\begin{equation}
\label{DefCirr}
  \mathbf{\chi'}_{irr,p}(\tau) = C_{irr,p}(\tau)
\mathbf{\chi}_{irr,p}(\tau) \; .
\end{equation}
With equation \eqref{DefTauS} one gets the
$\mathcal{S}$-transformation of $\mathbf{\chi'}_{irr,p}(\tau)$:
\begin{eqnarray*}
\mathbf{\chi'}_{irr,p}\left(-\frac{1}{\tau}\right) &=& {C_{irr,p}}\left(-\frac{1}{\tau}\right) \mathbf{\chi}_{irr,p}\left(-\frac{1}{\tau}\right) \\
&=& {C_{irr,p}}\left(-\frac{1}{\tau}\right) \boldsymbol{S}_p(\tau) {C_{irr,p}}^{-1}(\tau) {C_{irr,p}}(\tau) \mathbf{\chi}_{irr,p}(\tau) \\
&=& \underbrace{{C_{irr,p}}\left(-\frac{1}{\tau}\right)
\boldsymbol{S}_p(\tau) {C_{irr,p}}^{-1}(\tau)}_{=:\mathbf{S'}_p(\tau)}
\mathbf{\chi'}_{irr,p}(\tau) \; .
\end{eqnarray*}
$\mathbf{S'}_p(\tau)$ is now set to be equal to
$\mathsf{S}(p)$. So the matrix ${C_{irr,p}}(\tau)$ we are
looking for should relate $\mathsf{S}(p)$ and
$\boldsymbol{S}_p(\tau)$ through
\begin{equation}
\label{SCSCirr}
 \mathsf{S}(p) = {C_{irr,p}}\left(-\frac{1}{\tau}\right) \boldsymbol{S}_p(\tau)
{C_{irr,p}}^{-1}(\tau) \; .
\end{equation}
The detailed calculations leading to $C_{irr,p}(\tau)$  are contained in the thesis of HK \protect\cite{Knuth:2006}.\\
Here we just state the result, which we have verified calculating $\mathsf{S}(p)$ block by block with eq. \eqref{SCSCirr} and  
\begin{eqnarray}
\nonumber {C_{irr,p}}(\tau)&=&\mathtt{1\!\!l}_{2 \times 2} \oplus
\bigoplus^{p-1}_{s=1} {(C_{irr,p})_s}(\tau)\; ,\\
 \label{Cirrblock}
 {(C_{irr,p})_s}(\tau)&=&\begin{pmatrix}
 \frac{s+p}{2 p}-\mathrm{i}\frac{p-s}{2 p}\tau &\frac{s}{2 p}+\mathrm{i}\frac{s}{2 p}\tau\\
 \frac{p-s}{2 p}+\mathrm{i}\frac{p-s}{2 p}\tau &\frac{2 p -
s }{2 p}-\mathrm{i}\frac{s}{2 p}\tau
                      \end{pmatrix}\; .
\end{eqnarray}
This matrix only replaces the factor $j_p(\gamma,\tau)$ for the
case of $\gamma=\mathcal{S}$. Because $j_p(\gamma,\tau)$ depends
on $\gamma$, the matrix replacing it for other
$\gamma\neq\mathcal{S}$ is different from ${C_{irr,p}}(\tau)$.
Hence other elements of the representation $\rho(\gamma)$ are not
given by the transformation $\gamma$ of the same linear
combination of characters, $\mathbf{\chi'}_{irr,p}(\tau)$. In other words the
interpretation, it yields for $\mathsf{S}(p)$, does not hold
for the whole representation $\rho(\gamma)$. Consequently the
matrix ${C_{irr,p}}(\tau)$ is of little importance for the
original method of Fuchs et al.. It only gives us the new perspective explained just before eq. \eqref{DefCirr}. \\
However, for the extension of this method to indecomposable
representations this matrix is very helpful
to find the matrix $C_p$ connecting the larger S-matrix, $S_p$, taking
the place of $\mathsf{S}(p)$ to the $\alpha$-dependent
S-matrix $\mathtt{S}_{p,\alpha}$ from section \ref{SecFlohr}.
We have seen in section \ref{SecFlohr} that
$\mathtt{S}_{p,\alpha}$ belongs to a $SL(2,\mathbb{Z})$
representation $\mathtt{G}_{p,\alpha}(\gamma)$ (eq.
\ref{GalphaHom}). This representation gives the modular
transformation properties of a set of forms
$\pmb{\chi}_p(\alpha)(\tau)$ without any automorphy factor (eq.
\ref{GalphaRep}). So we get with the product $C_p
\mathtt{G}_{p,\alpha}(\gamma) {C_p}^{-1}$ another representation
of the modular group, which also needs no automorphy factor -- or
said in another way, its automorphy factor is the unit matrix.
Thus we can interpret this new representation as the one,
which gives directly the modular transformation properties of the
set of linear combinations $C_p \pmb{\chi}_p(\alpha)(\tau)$.

\subsubsection{Substitution of $\mathbf{\tau}$-Dependent Linear Combinations}
\label{substaualpha} We now start to compare the two methods
described in this section and section \ref{SecFlohr}. Some
character identities will help to transfer the $\tau$-dependent
matrices $\boldsymbol{S}_p(\tau)$ and $C_{irr,p}(\tau)$ into
$\alpha$-dependent matrices. This will reveal the connection
between $\boldsymbol{S}_p(\tau)$ and $\mathtt{S}_{p,\alpha}$. We will also use the $\alpha$-dependent pendant of
$C_{irr,p}(\tau)$, which we call $C'_{p}(\alpha)$, to find $C_p$ later on.
\liketheorem{Lemma:}{} The characters $\chi^+_{p-s,p}$ and $\chi^-_{p-s,p}$ given by the matrix elements of $B$ in eq. \eqref{Belements} (the 2nd and 3rd row, respectively) and the forms from eq. \eqref{ChiTildeDef}
fulfil the equation
\begin{equation} \label{charident}\mathrm{i} (s-p) \tau \chi^+_{p-s,p} + \mathrm{i} s \tau
\chi^-_{p-s,p} = - \frac{1}{\alpha} {\tilde{\chi}}_{p-s,p}(\alpha)
+ \frac{2}{\alpha} \chi^+_{p-s,p} + \frac{2}{\alpha}
\chi^-_{p-s,p} \; .
\end{equation}\qef
\likeremark{Remark:}{} For the two matrices $\boldsymbol{S}_p(\tau)$ (eq. \eqref{DefTauS} with $\gamma=\mathcal{S}$) and $C_{irr,p}(\tau)$ (eq. \eqref{Cirrblock}) one observes the following: The
only difference in the parts linear in $\tau$ of the pairs of matrix elements in the same row, of which one is multiplied by $\chi^+_{p-s,p}$ and the other with $\chi^-_{p-s,p}$ in eqns. \eqref{DefTauS} and \eqref{DefCirr}, is a factor of $(s-p)$ in the first and $s$ in the second matrix element. So for $0<s<p$ the constellation
given on the left hand side of equation \eqref{charident} appears
in the $\tau$-dependent linear combination of characters all the
time. We want to replace this by the right hand side using $2p
\times (3p-1)$ matrices, which are multiplied now by the vector
$\pmb{\chi}_p(\alpha)$ (eq. \eqref{charvecalpha}) instead of $\pmb{\chi}_{irr,p}\left(\tau\right)$ , but give the same result. \qef
\proof{$\!\!\!\!\!\!$:} We plug in the characters from equations
\eqref{Bdef} and \eqref{Belements} and find that the factors match
in precisely the way to let the dependence on $\Theta_{p-s,p}$ and
on $s$ drop out.
\begin{eqnarray*}
&& \mathrm{i}(s-p) \tau \biggl( \frac{1}{p \eta} \left[ s \Theta_{p-s,p} + (\partial \Theta)_{p-s,p} \right] \biggr) + \mathrm{i} s \tau \biggl( \frac{1}{p \eta} \left[ (p-s) \Theta_{p-s,p} - (\partial \Theta)_{p-s,p} \right] \biggr) \\
 &=& - \mathrm{i} \tau \frac{1}{\eta} (\partial \Theta)_{p-s,p} = - \frac{1}{\eta}(\nabla
 \Theta)_{p-s,p}\; .
\end{eqnarray*}
Equation \eqref{ChiTildeDef} guides the way to insert a zero (one
of two we need to insert here):
\begin{eqnarray}
\nonumber && - \frac{\alpha}{\alpha \eta}(\nabla \Theta)_{p-s,p} -
\frac{1}{\alpha \eta} 2\Theta_{p-s,p} + \frac{1}{\alpha \eta}
2\Theta_{p-s,p}\\ \nonumber
 &=& - \frac{1}{\alpha} {\tilde{\chi}}_{p-s,p}(\alpha) + \frac{2}{\alpha} \frac{1}{p\eta}\, s\, \Theta_{p-s,p} + \frac{2}{\alpha} \frac{1}{p\eta} (\partial \Theta)_{p-s,p} + \frac{2}{\alpha} \frac{1}{p\eta} (p-s) \Theta_{p-s,p} - \frac{2}{\alpha} \frac{1}{p\eta} (\partial \Theta)_{p-s,p} \\ \nonumber
 &=& - \frac{1}{\alpha} {\tilde{\chi}}_{p-s,p}(\alpha) + \frac{2}{\alpha} \chi^+_{p-s,p} + \frac{2}{\alpha}
 \chi^-_{p-s,p} \; .
\end{eqnarray}\qed
We start with the matrix $\boldsymbol{S}_p(\tau)$ and write down its
partner $2p \times (3p-1)$ matrix . A column must be inserted for
each form ${\tilde{\chi}}_{s,p}(\alpha)$ after the columns
multiplied by $\chi^+_{s,p}$ and $\chi^-_{s,p}$ for $0<s<p$. In
the elements in the latter two columns the respective factors
$\mathrm{i}(s-p)\tau$ and $\mathrm{i} s \tau$ are both replaced by $2/\alpha$. The
added column has to contain $-1/\alpha$. This way we do the
following changes for the blocks of $\boldsymbol{S}_p(\tau)$:
\begin{equation*}
  \begin{pmatrix}
 \frac{s}{p}\mathfrak{c}_{sl}-\mathrm{i}\tau \frac{p-j}{p}\mathfrak{s}_{sl} & \frac{s}{p}\mathfrak{c}_{sl}+\mathrm{i}\tau \frac{j}{p}\mathfrak{s}_{sl}\\
 \frac{p-s}{p}\mathfrak{c}_{sl}+\mathrm{i}\tau
\frac{p-j}{p}\mathfrak{s}_{sl} &
\frac{p-s}{p}\mathfrak{c}_{sl}-\mathrm{i}\tau \frac{j}{p}\mathfrak{s}_{sl}
  \end{pmatrix} \rightarrow
  \begin{pmatrix}
 \frac{s}{p} \mathfrak{c}_{sl} + \frac{2}{p} \frac{1}{\alpha} \mathfrak{s}_{sl}& \frac{s}{p} \mathfrak{c}_{sl} + \frac{2}{p} \frac{1}{\alpha} \mathfrak{s}_{sl} & -\frac{1}{p \alpha} \mathfrak{s}_{sl} \\
 \frac{p-s}{p} \mathfrak{c}_{sl} - \frac{2}{p}
\frac{1}{\alpha} \mathfrak{s}_{sl} & \frac{p-s}{p}
\mathfrak{c}_{sl} - \frac{2}{p} \frac{1}{\alpha} \mathfrak{s}_{sl}
& \frac{1}{p \alpha} \mathfrak{s}_{sl}
 \end{pmatrix}\; .
\end{equation*}
The first two rows of the added columns are zero because these
rows do not depend on $\tau$. We see that the matrix we get is
just a composition of the $\alpha$-dependent S-matrix
,$\mathtt{S}_{p,\alpha}$, and a subsequent projection onto the
components of $\pmb{\chi}_p(\alpha)$ belonging to irreducible
representations, as it is expected to be (cf. \eqref{FlohrS}).\\
More interesting is the application to the matrix
$C_{irr,p}(\tau)$. $2 \times 2$ blocks on the diagonal get replaced by $2 \times
3$ blocks arranged in a diagonal way, i.e. the whole matrix,
called $C'_p$, is the direct sum of a $2 \times 2$ unit matrix and
these blocks.
\begin{eqnarray}
\nonumber \begin{pmatrix}
 \frac{s+p}{2 p}-\mathrm{i}\frac{p-s}{2 p}\tau &\frac{s}{2 p}+\mathrm{i}\frac{s}{2 p}\tau\\
 \frac{p-s}{2 p}+\mathrm{i}\frac{p-s}{2 p}\tau &\frac{2 p -
s }{2 p}-\mathrm{i}\frac{s}{2 p}\tau
                      \end{pmatrix}\rightarrow \underbrace{\begin{pmatrix}
 \frac{s+p}{2 p}+\frac{1}{2 p \alpha} &\frac{s}{2 p}+\frac{1}{2 p \alpha} & -\frac{1}{2 p \alpha}\\
 \frac{p-s}{2 p}-\frac{1}{2 p \alpha} &\frac{2 p -
s }{2 p}-\frac{1}{2 p \alpha} &\frac{1}{2 p \alpha}
                      \end{pmatrix}}_{(C'_{p})_s(\alpha)}\quad ,\\
                \label{Cprimealpha}
C'_{p}(\alpha)=\mathtt{1\!\!l}_{2 \times 2} \oplus
\bigoplus_{s=1}^{p-1} (C'_{p})_s(\alpha)\; .
\end{eqnarray}
Now the matrix $C'_{p}(\alpha)$ encodes the $\tau$-dependent linear combinations of
characters given by $C_{irr,p}$, for which $\mathsf{S}_{p}$ gives their $\mathcal{S}$-transformation, as $\tau$-independent linear
combinations of these characters and the forms
${\tilde{\chi}}_{s,p}(\alpha)$. Furthermore we use these linear combinations also in our extended block diagonalisation method, as we demand the matrices $C_p$ to contain the matrix $C'_{p}(\alpha)$  in the rows corresponding to the irreducible representations.

\subsubsection{Calculation of $P_p$}
For the calculation of $P_p$ we need the matrices $S_p$ and $K_p$.
From the considerations for $p=2$ and $p=3$ in appendix \ref{SCKfromp2} we can directly do the step to arbitrary $p$. The following generalisation from $C_2(\alpha)$ (eq. \ref{C2expl}) and $C_3(\alpha)$ (eq. \ref{C3expl}) suggests itself:
\begin{eqnarray}
 \label{Cpexpl} {C_{p}}(\alpha)&=& \boldsymbol{\mathtt{1\!\!l}}_{2\times2}
  \oplus \bigoplus_{s=1}^{p-1} {C_{p,s}}(\alpha)\; ,\\
\nonumber  {C_{p,s}}(\alpha) &=& \begin{pmatrix}
  \frac{p+s}{2p}+ \frac{1}{p\alpha} & \frac{s}{2p}+ \frac{1}{p\alpha} & -\frac{1}{2p\alpha}\\
  \frac{p-s}{2p}- \frac{1}{p\alpha} & \frac{2p-s}{2p}- \frac{1}{p\alpha} & \frac{1}{2p\alpha} \\
  \frac{p+s}{p}+ \frac{2}{p\alpha} & \frac{2p+s}{p}+ \frac{2}{p\alpha} & -\frac{1}{p\alpha}
  \end{pmatrix}\; .
\end{eqnarray}
Its inverse is
\begin{eqnarray*}
  {C_{p}}^{-1}(\alpha)&=& \boldsymbol{\mathtt{1\!\!l}}_{2\times2}
  \oplus \bigoplus_{s=1}^{p-1} {C_{p,s}}^{-1}(\alpha)\; ,\\
 {C_{p,s}}^{-1}(\alpha) &=& \begin{pmatrix}
  2 & 1 & -\frac{1}{2}\\
  -1 & 0 & \frac{1}{2} \\
  s\alpha+2 &(p+s)\alpha+2 & -\frac{p}{2}\alpha
  \end{pmatrix}\; .\\
\end{eqnarray*}
We now calculate the $\alpha$-independent S-matrix, $S_p$, block by block using eq. \eqref{SCalphaSCinvalpha}. With the blocks of $\mathtt{S}_{p,\alpha}$ (eq.
\eqref{FlohrS}) we have to determine the following expressions:
\begin{eqnarray*}
& S(p)_{0,l}{C_{p,l}}^{-1}(\alpha)\; ,&\\
& {C_{p,s}}(\alpha) S(p)_{s,0}\; ,&\\
& {C_{p,s}}(\alpha) S(p,\alpha)_{s,l}{C_{p,l}}^{-1}(\alpha)\; .&
\end{eqnarray*}
The block $S(p)_{0,0}$ is not touched at all. $S(p)_{0,l}$ and
$S(p)_{s,0}$, for $0<s,l<p$, are also not changed by the
multiplication. And the last product gives
\begin{eqnarray}
\label{SExtBlock}
&&S(p)_{s,l}={C_{p,s}}(\alpha) S(p,\alpha)_{s,l}{C_{p,l}}^{-1}(\alpha)=\frac{2}{\sqrt{2p}}(-1)^{p+l+s}\times \\
&&\begin{pmatrix}   \frac{s}{p}\cos{(\pi
\frac{sl}{p})}+\frac{p-l}{p}\sin{(\pi \frac{sl}{p})} &
\frac{s}{p}\cos{(\pi \frac{sl}{p})}- \frac{l}{p}\sin{(\pi
\frac{sl}{p})} & 0\\
  \nonumber \frac{p-s}{p}\cos{(\pi
\frac{sl}{p})}-\frac{p-l}{p}\sin{(\pi \frac{sl}{p})} &
\frac{p-s}{p}\cos{(\pi \frac{sl}{p})}+ \frac{l}{p}\sin{(\pi
\frac{sl}{p})} & 0 \\
2\cos{(\pi \frac{sl}{p})}+2\sin{(\pi \frac{sl}{p})} & 2\cos{(\pi
\frac{sl}{p})}+ 2\sin{(\pi \frac{sl}{p})} & -\sin{(\pi
\frac{sl}{p})}
\end{pmatrix}\; .
\end{eqnarray}
We plug the matrix elements of $S_p$ into our result for $K_p$ in eq. \eqref{KExtDef} (we have
set ${k^{(s)}}_{1}={k^{(s)}}_{2}=0$):
\begin{eqnarray}
\label{Kexplp}
  K_p&=& K_{p,I} \oplus \bigoplus_{l=1}^{p-1} K_{p,l}\; ,\\
\nonumber K_{p,0}&=&\begin{pmatrix}
     \sqrt{2p^3} & 0 \\
     0 & (-1)^{p+1}\sqrt{2p^3}
     \end{pmatrix}\; ,\\
\nonumber K_{p,l}&=&\begin{pmatrix}
     (-1)^{p+l+1}\sqrt{\frac{p}{2}}\,\mathfrak{s}_{1l} & (-1)^{p+l}\sqrt{\frac{2}{p^3}}\left(\mathfrak{c}_{1l} - l \mathfrak{s}_{1l}\right) & 0\\
     (-1)^{p+l}\sqrt{\frac{p}{2}}\,\mathfrak{s}_{1l} & (-1)^{p+l+1}\sqrt{\frac{2}{p^3}}\left(\mathfrak{c}_{1l} + (p-l) \mathfrak{s}_{1l}\right) & 0\\
    0 & 0 & 1
   \end{pmatrix}\; .
\end{eqnarray}
The matrices $K_p$ and $S_p$ determine $P_p$ through equation \eqref{PSK}. Explicitly we get for $P_p$:
\begin{eqnarray}
\label{Pexpl}
 P_p&=&\begin{pmatrix}
P(p)_{0,0} & P(p)_{0,1} & \dotso & P(p)_{0,p-1}  \\
P(p)_{1,0} & P(p)_{1,1} & \dotso & P(p)_{1,p-1} \\
\vdots & \vdots & \ddots & \vdots &\\
P(p)_{p-1,0} & P(p)_{p-1,1} & \dotso & P(p)_{p-1,p-1}
\end{pmatrix}\; , \vspace{5mm}\\
\nonumber P(p)_{0,0}&=&\frac{1}{\sqrt{2p}}
\begin{pmatrix}
p & (-1)^{p+1}p \\
p & -p
                           \end{pmatrix}\; , \vspace{5mm}\\
 \nonumber P(p)_{0,l}&=&\frac{2}{\sqrt{2p}}\begin{pmatrix}
0 & (-1)^{p+l+1}\frac{2}{p}\mathfrak{s}_{1l} &0\\
0 & -\frac{2}{p}\mathfrak{s}_{1l} &0
                           \end{pmatrix} \vspace{5mm}\\
 \nonumber P(p)_{s,0}&=&\frac{1}{\sqrt{2p}}\begin{pmatrix}
 s & (-1)^{s+1}s \\
 p-s & (-1)^{s+1}(p-s)\\
 2p & 2(-1)^{s+1}p
\end{pmatrix}\; , \vspace{5mm}\\
\nonumber P(p,\alpha)_{s,l}&=& \begin{pmatrix}

(-1)^{s+1}\frac{\mathfrak{s}_{sl}}{\mathfrak{s}_{1l}}& (-1)^{s+1}
\frac{2}{p^2} \left(s \mathfrak{c}_{sl} \mathfrak{s}_{1l} -
\mathfrak{s}_{sl} \mathfrak{c}_{1l} \right) & 0 \\
(-1)^{s}\frac{\mathfrak{s}_{sl}}{\mathfrak{s}_{1l}}& (-1)^{s+1}
\frac{2}{p^2} \left((p-s) \mathfrak{c}_{sl} \mathfrak{s}_{1l} +
\mathfrak{s}_{sl} \mathfrak{c}_{1l} \right) & 0 \\
 0 & (-1)^{s+1} \frac{4}{p} \left(
\mathfrak{c}_{sl}  + \mathfrak{s}_{sl} \right)\mathfrak{s}_{1l} &
(-1)^{p+s+l+1}\sqrt{\frac{2}{p}}\mathfrak{s}_{sl}
 \end{pmatrix}\; .
\end{eqnarray}
If we plug equation \eqref{PSK} into eq. \eqref{NPMP} the \mbox{pre-fusion} coefficients are given by the ''generalised'' Verlinde formula:
\begin{equation}
\label{NextSMS}
 N_{p,I} = {S}_{p} K_{p} M_{p,I} (K_{p})^{-1}
 {S}_{p}\; .
\end{equation}
We have now all ingredients to carry through calculations for any
value of $p$ in our extension of the method of Fuchs et. al..
$P_p$ is also in the general case invertible because with the
invertible matrix $\mathtt{S}_{p,\alpha}$ also  $S_p$ has to be
invertible and $K_p$ has been constructed as a full rank matrix.

\section{Projection of The Extended Block Diagonalisation Method on Irreducible Representations}\label{Projection}
This extension to indecomposable representations reduces in every step by simple projection on the first two rows and columns of every $2 \times 3$, $3 \times 2$ and $3 \times 3$ block to the original work of Fuchs et al.. This is also shown in this section in line with the proof that the results for the pre-fusion rules of irreducible times irreducible representations are the same in the small and the extended version. 
For this task we change the sequence of the representations
from the groups of three -- two irreducible and one indecomposable
representations -- to the following one:
\begin{eqnarray}
\label{newseq} &&\left[ h_{1,p}\right],\;\left[
h_{1,2p}\right],\;\left[ h_{1,1}\right],\;\left[
h_{1,2p+1}\right],\;\left[ h_{1,2}\right],\;\left[
h_{1,2p+2}\right], \ldots,\;\left[ h_{1,p-1}\right],\;\left[
h_{1,3p-1}\right],\\ \nonumber &&\left[
\widetilde{h_{1,p+1}}\right],\;\left[
\widetilde{h_{1,p+2}}\right],\ldots,\;\left[
\widetilde{h_{1,2p-1}}\right]
\end{eqnarray}
with the indecomposable representations all put to the end. This
leads to the permutation of both rows and columns in the
matrices $S_p$, $K_p$, $P_p$, $M_{p,I}$ and finally $N_{p,I}$.
Also the sequence of the latter two groups of matrices is changed,
as the index $I$ is affected by the same permutation. The reason
is the form all these matrices take after the permutation. All the
zeros, which we inserted in some matrices and consequently
appeared in other matrices are grouped together with the
indecomposable
representations in the last columns. \\
We introduce the following notation, which tells us that a matrix
has some form without specifying all matrix elements or the size
of the matrix. The matrix $S_p$ has now the form (cf. eq.
\eqref{SExtBlock}):
\begin{equation*}
 S_p = \begin{pmatrix}
 \setlength\fboxsep{0.7cm}\fbox{\makebox[0.4cm]{$\mathsf{S}(p)$}}
\makebox[1cm]{0}\\
\setlength\fboxsep{0.5cm}\fbox{\makebox[1.8cm]{ }}
 \end{pmatrix}\; .
\end{equation*}
This states that the box on the upper left contains exactly the
matrix $\mathsf{S}(p)$ (eq. \eqref{SirrDef}), the box at the bottom
contains the other a priori non-zero elements of $S_p$ and on the
upper right all matrix elements are zero. With this notation we
give the statement, which we want to proof.
\liketheorem{Proposition:}{} The pre-fusion coefficients matrices,
$N_{p,I}$, each contain the coefficients of the ''small'' pre-fusion
algebra, $N_{irr,p, I}$, calculated in \protect\cite{Fuchs:2003yu} for $0 < I \leq 2p$ in the following form:
\begin{equation}
\label{NpIform}
  N_{p,I}= \begin{pmatrix}
 \setlength\fboxsep{0.7cm}\fbox{\makebox[0.4cm]{$N_{irr,p, I}$}}
\makebox[1cm]{$0$}\\
\setlength\fboxsep{0.5cm}\fbox{\makebox[1.8cm]{ }}
 \end{pmatrix}\; .
\end{equation}\qef
\likeremark{Remark:}{} We call the matrices, which appear in the ''small'' block diagonalisation method (cf. \protect\cite{Fuchs:2003yu}) and correspond to $K_p$, $P_p$, $M_{p,I}$ and $N_{p,I}$ as defined for the extended one, $K_{irr,p}$, $P_{irr,p}$, $M_{irr,p,I}$ and $N_{irr,p,I}$, respectively. \qef
\proof{$\!\!\!\!\!\!$:} The only coefficients of the matrix $K_p$,
which are different from zero and do not come from the matrix
$K_{irr,p}$, are the additional diagonal matrix elements. The
permutation of rows and columns leaves them on the diagonal and
assembles them in a block, which is equal to the unit matrix in
$p-1$ dimensions (cf. eq. \eqref{KExtDef}):
\begin{equation*}
  K_p= \begin{pmatrix}
 \setlength\fboxsep{0.7cm}\fbox{\makebox[0.4cm]{$K_{irr,p}$}}
\makebox[1cm]{0}\\
\makebox[1.85cm]{0}\setlength\fboxsep{0.5cm}\fbox{\makebox[0cm]{$\mathtt{1\!\!l}$}}
 \end{pmatrix}\; .
\end{equation*}
The matrix $P$ consequently looks like
\begin{equation}
\label{Ppform} P_p= S_p K_p = \begin{pmatrix}
 \setlength\fboxsep{0.7cm}\fbox{\makebox[0.4cm]{$P_{irr,p}$}}
\makebox[1cm]{0}\\
\setlength\fboxsep{0.5cm}\fbox{\makebox[1.8cm]{ }}
 \end{pmatrix}\; .
\end{equation}
Furthermore we need the form of the inverse of $P_p$. As $(K_p)^{-1}$ is the direct sum of  $(K_{irr,p})^{-1}$ and the unit matrix, $\boldsymbol{\mathtt{1\!\!l}}_{(p-1)\times(p-1)}$, we have
\begin{equation}
\label{Ppinvform} (P_p)^{-1} =(K_p)^{-1} S_p = \begin{pmatrix}
 \setlength\fboxsep{0.7cm}\fbox{\makebox[0.4cm]{$(P_{irr,p})^{-1}$}}
\makebox[1cm]{0}\\
\setlength\fboxsep{0.5cm}\fbox{\makebox[1.8cm]{ }}
 \end{pmatrix}\; .
\end{equation}
We also note, that the last $p-1$ rows of this matrix are equal to those of the S-matrix, $S_p$.\\
We construct the matrices $M_{p,I}$ for $0 < I \leq 2p$ in the new
sequence. For each block defined in equation \eqref{defMp} (in the
sequence of representations we used there) the element
$(M_{p,I,n})_{13}=(P_{p})_I^{3n+2}$ is zero (cf. eq. \eqref{Pexpl}). These zeros are of
interest because the permutation to the new sequence of
representations bring them from the 5th, 8th, 11th etc. column,
where they are not on the diagonal, to a new position in the last
$p-1$ columns and the first $2p$ rows, which need to be zero, as
we will see next. Hence $M_{p,I}$ appears in the form
\begin{equation}
\label{MpIform} M_{p,I}= \begin{pmatrix}
 \setlength\fboxsep{0.7cm}\fbox{\makebox[0.4cm]{$M_{irr,p,I}$}}
\makebox[1cm]{0}\\
\makebox[1.85cm]{0}\setlength\fboxsep{0.5cm}\fbox{}
 \end{pmatrix}\; .
\end{equation}
We end up with the product (see eqns. \eqref{Ppform},
\eqref{MpIform} and \eqref{Ppinvform}) for $0 < I \leq 2p$
\begin{eqnarray*}
  N_{p,I} &=& P_p M_{p,I} (P_p)^{-1} \\ 
  &=& \begin{pmatrix}
 \setlength\fboxsep{0.7cm}\fbox{\makebox[0.4cm]{$P_{irr,p}$}}
\makebox[1cm]{0}\\
\setlength\fboxsep{0.5cm}\fbox{\makebox[1.8cm]{ }}
 \end{pmatrix}\begin{pmatrix}
 \setlength\fboxsep{0.7cm}\fbox{\makebox[0.4cm]{$M_{irr,p,I}$}}
\makebox[1cm]{0}\\
\makebox[1.85cm]{0}\setlength\fboxsep{0.5cm}\fbox{}
 \end{pmatrix}\begin{pmatrix}
 \setlength\fboxsep{0.7cm}\fbox{\makebox[0.4cm]{$(P_{irr,p})^{-1}$}}
\makebox[1cm]{0}\\
\setlength\fboxsep{0.5cm}\fbox{\makebox[1.8cm]{ }}
 \end{pmatrix}\;,
\end{eqnarray*}
which has the form given in eq. \eqref{NpIform}.\qed

\section{Equivalence of Both Approaches}
 \label{SecEqui}
The limit-Verlinde formula, which we have learned about in
section \ref{SecFlohr}, expresses the possibility to
simultaneously diagonalise the set of matrices
$\mathtt{N}_{p,I}(\alpha)$. Unfortunately these are not the
matrices of pre-fusion coefficients as in the case of rational
conformal field theories. They rather only become matrices of
pre-fusion coefficients after the limit $\alpha \rightarrow 0$ has
been taken -- to be precise we can map these pre-fusion coefficients then to
the proposed true fusion coefficients in an unambiguous way. But
still it gives us the possibility to write the equation for the
matrix elements of $\mathtt{N}_{p,I}(\alpha)$ \eqref{Nijkalpha} as
\begin{equation}
\label{NalphaSMS} \mathtt{N}_{p,I}(\alpha)=
\mathtt{S}_{p,\alpha} M_{diag,p,\alpha,I}
 {\mathtt{S}_{p,\alpha}}^{-1}\; ,
\end{equation}
with $M_{diag,p,\alpha,I}$ given by
\begin{equation*}
  M_{diag,p,\alpha,I} = diag \left( \frac{{(\mathtt{S}_{p,\alpha})_{I}}^1}{{(\mathtt{S}_{p,\alpha})_{3}}^1}, \frac{{(\mathtt{S}_{p,\alpha})_{I}}^2}{{(\mathtt{S}_{p,\alpha})_{3}}^2}, \ldots,
 \frac{{(\mathtt{S}_{p,\alpha})_{I}}^{3p-1}}{{(\mathtt{S}_{p,\alpha})_{3}}^{3p-1}}\right)\; ,
\end{equation*}
One can also introduce the matrix $K_{diag,p,\alpha}$ defined as the
diagonal matrix with the reciprocal value of the elements of the vacuum row of $\mathtt{S}_{p,\alpha}$ on the diagonal,
\begin{equation}
\label{Kdiagalphadef}
 K_{diag,p,\alpha}=diag \left( \frac{1}{{(\mathtt{S}_{p,\alpha})_{3}}^1}, \frac{1}{{(\mathtt{S}_{p,\alpha})_{3}}^2}, \ldots,
 \frac{1}{{(\mathtt{S}_{p,\alpha})_{3}}^{3p-1}}\right)\; ,
\end{equation}
which of course commutes in equation \eqref{NalphaSMS} with the
matrices $M_{diag,p,\alpha,I}$, because these are also diagonal. This way we are able to see it parallel to our earlier notation.
$M_{diag,p,\alpha,I}$ is given by the I-th row of the product
$\mathtt{S}_{p,\alpha}K_{diag,p,\alpha}$ and
\begin{equation*}
\mathtt{N}_{p,I}(\alpha)= \mathtt{S}_{p,\alpha} K_{diag,p,\alpha}
M_{diag,p,\alpha,I} {K_{diag,p,\alpha}}^{-1}
 {\mathtt{S}_{p,\alpha}}^{-1}\; .
\end{equation*}
This gives a more rounded picture and helps us to show the central theorem of this paper.
\liketheorem{Theorem:}{} The pre-fusion coefficients calculated
with the limit-Verlinde formula are the same as the ones
calculated with the extended block diagonalisation method:
\begin{equation}
\label{EquiStat}
 \lim_{\alpha\rightarrow 0} {\mathtt{N}_{ij}}^k(\alpha) =
 {(N_{p,I})_j}^k\; .
\end{equation}
\qef \proof{$\!\!\!\!\!\!$:} We plug equations
\eqref{NalphaSMS}and \eqref{NextSMS} into eq. \eqref{EquiStat} and
have
\begin{equation}
\label{EquiProp} \lim_{\alpha\rightarrow 0}
\left(\mathtt{S}_{p,\alpha} M_{diag,p,\alpha,I}
{\mathtt{S}_{p,\alpha}} \right) = S_p K_p M_{p,I} (K_p)^{-1} S_p
\; .
\end{equation}
We insert two unit matrices on the left hand side of this
equation:
\begin{equation}
\nonumber \mathtt{S}_{p,\alpha} M_{diag,p,\alpha,I}
\mathtt{S}_{p,\alpha} = \mathtt{S}_{p,\alpha} E_{p,\alpha}
{E_{p,\alpha}}^{-1} M_{diag,p,\alpha,I} E_{p,\alpha}
{E_{p,\alpha}}^{-1} \mathtt{S}_{p,\alpha}
\end{equation}
with $E_{p,\alpha}$ defined as
\begin{equation} \label{Edef}
E_{p,\alpha}:={\mathtt{S}_{p,\alpha}}^{-1} S_p =
{\mathtt{S}_{p,\alpha}} S_p\; ,
\end{equation}
in order to have
\begin{equation} \label{SEMES} \mathtt{S}_{p,\alpha} M_{diag,p,\alpha,I}
\mathtt{S}_{p,\alpha} = S_p {E_{p,\alpha}}^{-1}
M_{diag,p,\alpha,I} E_{p,\alpha} S_p\; ,
\end{equation}
With a block diagonal ansatz one can directly
calculate the blocks of $E_{p,\alpha}$ with equation \eqref{Edef}:
\begin{eqnarray*}
 E_{p,\alpha}&=&\boldsymbol{\mathtt{1\!\!l}}_{2\times2}
  \oplus \bigoplus_{s=1}^{p-1} (E_{p,\alpha})_s\\
  (E_{p,\alpha})_s&=& S(p,\alpha)_{s,s} S(p)_{s,s} = \begin{pmatrix}
   \frac{s}{p} - \frac{2}{p \alpha} &    \frac{s}{p} - \frac{2}{p
   \alpha} & \frac{1}{p \alpha}\\
   \frac{p-s}{p} + \frac{2}{p \alpha} &    \frac{p-s}{p} + \frac{2}{p
   \alpha} & -\frac{1}{p \alpha}\\
   2-(p-s)\alpha & 2+s\alpha & 0
  \end{pmatrix}\; .
\end{eqnarray*}
where we used the blocks from eqns. \eqref{FlohrS} and
\eqref{SExtBlock}. \\ We are going to show that the product
${E_{p,\alpha}}^{-1} M_{diag,p,\alpha,I} E_{p,\alpha} $ has a well
defined limit for $\alpha \rightarrow 0$. This is not clear. For
the whole term on the right hand side of equation \eqref{SEMES} this limit is
well defined. They are the coefficients
${\mathtt{N}_{ij}}^k(\alpha)$. But still singular terms in the
mentioned product could drop out through the multiplication of
$S_p$ from both
sides.\\
We simply calculate first the matrices $M_{diag,p,\alpha,I}$. We
need to consider the following cases. For $I=1,2$ the matrices
$M_{diag,p,\alpha,I}$ differ by minus signs. There are three
more groups to be distinguished, which belong each to one row of
the $3 \times 3$ blocks of $\mathtt{S}_{p,\alpha}$. We use again the same
abbreviations as for $\mathtt{S}_{p,\alpha}$ in eq.
\eqref{FlohrS}.
\begin{eqnarray}
 && M_{diag,p,\alpha,I} = (M_{diag,p,\alpha,I})_0
  \oplus \bigoplus_{l=1}^{p-1} (M_{diag,p,\alpha,I})_l\\
\nonumber I=1,2:&& (M_{diag,p,\alpha,I})_0=\begin{pmatrix} p & 0 \\
                                                0 & (-1)^{Ip} p
                                                \end{pmatrix}\\
                                                \nonumber
      && (M_{diag,p,\alpha,I})_l=(-1)^{I(p-l)}\begin{pmatrix} \frac{-p \alpha}{\alpha \mathfrak{c}_{1l} + 2 \mathfrak{s}_{1l}} &0 & 0\\
0 & \frac{-p \alpha}{\alpha\mathfrak{c}_{1l} +
2 \mathfrak{s}_{1l}} & 0 \\
0 & 0 & 0
\end{pmatrix}\\
\nonumber I=3,6,\ldots:&& (M_{diag,p,\alpha,I})_0=\begin{pmatrix} I & 0 \\
                                                0 & (-1)^I I
                                                \end{pmatrix}\\
\nonumber      && (M_{diag,p,\alpha,I})_l=\begin{pmatrix}
(-1)^{I+1}\frac{I \alpha \mathfrak{c}_{Il} + 2
\mathfrak{s}_{Il}}{\alpha \mathfrak{c}_{1l} + 2
 \mathfrak{s}_{1l}} & 0 & 0 \\
0 & (-1)^{I+1}\frac{I \alpha \mathfrak{c}_{Il} + 2
\mathfrak{s}_{Il}}{\alpha \mathfrak{c}_{1l} + 2
 \mathfrak{s}_{1l}} & 0 \\
 0 & 0 & (-1)^{I+1}\frac{\mathfrak{s}_{Il}}{\mathfrak{s}_{1l}}
\end{pmatrix}\\
\nonumber I=4,7,\ldots:&&(M_{diag,p,\alpha,I})_0=\begin{pmatrix} p-I & 0 \\
                                                0 & (-1)^I (p-I)
                                                \end{pmatrix}\\
\nonumber     && (M_{diag,p,\alpha,I})_l=\begin{pmatrix}
(-1)^{I+1}\frac{(p-I) \alpha \mathfrak{c}_{Il} - 2
\mathfrak{s}_{Il}}{\alpha \mathfrak{c}_{1l} + 2
 \mathfrak{s}_{1l}} & 0 & 0 \\
0 & (-1)^{I+1}\frac{(p-I) \alpha \mathfrak{c}_{Il} - 2
\mathfrak{s}_{Il}}{\alpha \mathfrak{c}_{1l} + 2
 \mathfrak{s}_{1l}} & 0 \\
 0 & 0 & (-1)^{I}\frac{\mathfrak{s}_{Il}}{\mathfrak{s}_{1l}}
\end{pmatrix}\\
\nonumber I=5,8,\ldots:&&(M_{diag,p,\alpha,I})_0=\begin{pmatrix} 2p & 0 \\
                                                0 & (-1)^I 2p
                                                \end{pmatrix}\\
     \nonumber && (M_{diag,p,\alpha,I})_l=\begin{pmatrix} (-1)^{I} p \alpha \frac{(p-l) \alpha \mathfrak{s}_{Il} - 2
\mathfrak{c}_{Il}}{\alpha \mathfrak{c}_{1l} + 2
 \mathfrak{s}_{1l}} & 0 & 0 \\
0 & (-1)^{I+1} p \alpha \frac{l \alpha \mathfrak{s}_{sl} + 2
\mathfrak{c}_{Il}}{\alpha \mathfrak{c}_{1l} + 2
 \mathfrak{s}_{1l}} & 0 \\
 0 & 0 & 0
\end{pmatrix} \; .
\end{eqnarray}
For these four cases we can now calculate the product
\begin{eqnarray}
\label{MItilde} &&\tilde{M}_{\alpha,I}={E_{p,\alpha}}^{-1}
M_{diag,p,\alpha,I} E_{p,\alpha}= (\tilde{M}_{\alpha,I})_0
  \oplus \bigoplus_{l=1}^{p-1} \left[(-1)^{I}(\tilde{M}_{\alpha,I})_l\right] \\ \vspace{0.5cm}
\nonumber &&(\tilde{M}_{\alpha,I})_0 = (M_{diag,p,\alpha,I})_0  \\
\nonumber I&=&1,2:\\
\nonumber &&(\tilde{M}_{\alpha,I})_l = (-1)^{(p+l)}\begin{pmatrix}
-\frac{ l_1 }{\alpha \mathfrak{c}_{1l} + 2 \mathfrak{s}_{1l}} &
 -\frac{ l_1 }{\alpha
\mathfrak{c}_{1l} + 2 \mathfrak{s}_{1l}} & 0 \\
\frac{ l_2 }{\alpha \mathfrak{c}_{1l} + 2 \mathfrak{s}_{1l}}  &
\frac{ l_1 }{\alpha \mathfrak{c}_{1l} + 2 \mathfrak{s}_{1l}}  & 0
\\ 0 & 0 & -\frac{ \alpha p }{\alpha
\mathfrak{c}_{1l} + 2 \mathfrak{s}_{1l}}
\end{pmatrix} \vspace{0.2cm}\\
\nonumber I&=&3,6,9,\ldots:\\
\nonumber &&(\tilde{M}_{\alpha,I})_l = \begin{pmatrix} -\frac{
\left(I l_1 \mathfrak{c}_{Il} +
2p\mathfrak{s}_{Il}\right)\mathfrak{s}_{1l} - l_2
\mathfrak{c}_{1l} \mathfrak{s}_{Il}}{p \mathfrak{s}_{1l} \left(
\alpha \mathfrak{c}_{1l} + 2 \mathfrak{s}_{1l} \right)} &
-\frac{l_1 \left( I \mathfrak{s}_{1l} \mathfrak{c}_{Il} -
\mathfrak{c}_{1l} \mathfrak{s}_{Il}\right)}{p \mathfrak{s}_{1l}
\left( \alpha
\mathfrak{c}_{1l} + 2 \mathfrak{s}_{1l} \right)} & 0 \\
\frac{l_1 \left( I \mathfrak{s}_{1l} \mathfrak{c}_{Il} -
\mathfrak{c}_{1l} \mathfrak{s}_{Il}\right)}{p \mathfrak{s}_{1l}
\left( \alpha \mathfrak{c}_{1l} + 2 \mathfrak{s}_{1l} \right)} &
\frac{ \left(I l_2 \mathfrak{c}_{Il} -
2p\mathfrak{s}_{Il}\right)\mathfrak{s}_{1l} - l_1
\mathfrak{c}_{1l} \mathfrak{s}_{Il}}{p \mathfrak{s}_{1l} \left(
\alpha \mathfrak{c}_{1l} + 2 \mathfrak{s}_{1l} \right)} & 0
\\ 0 & 0 & -\frac{ I \alpha \mathfrak{c}_{Il} + 2 \mathfrak{s}_{Il}}{\alpha \mathfrak{c}_{1l} + 2 \mathfrak{s}_{1l}}
\end{pmatrix} 
\end{eqnarray}
\begin{eqnarray}
\nonumber I&=&4,7,10,\ldots:\\
\nonumber &&(\tilde{M}_{\alpha,I})_l = \begin{pmatrix} -\frac{
\left((p-I)l_1 \mathfrak{c}_{Il} -
2p\mathfrak{s}_{Il}\right)\mathfrak{s}_{1l} + l_2
\mathfrak{c}_{1l} \mathfrak{s}_{Il}}{p \mathfrak{s}_{1l} \left(
\alpha \mathfrak{c}_{1l} + 2 \mathfrak{s}_{1l} \right)} &
-\frac{l_1 \left( (p-I) \mathfrak{s}_{1l} \mathfrak{c}_{Il} +
\mathfrak{c}_{1l} \mathfrak{s}_{Il}\right)}{p \mathfrak{s}_{1l}
\left( \alpha
\mathfrak{c}_{1l} + 2 \mathfrak{s}_{1l} \right)} & 0 \\
\frac{l_1 \left( (p-I) \mathfrak{s}_{1l} \mathfrak{c}_{Il} +
\mathfrak{c}_{1l} \mathfrak{s}_{Il}\right)}{p \mathfrak{s}_{1l}
\left( \alpha \mathfrak{c}_{1l} + 2 \mathfrak{s}_{1l} \right)} &
 \frac{\left((p-I)l_2
\mathfrak{c}_{Il} + 2p\mathfrak{s}_{Il}\right)\mathfrak{s}_{1l} +
l_1 \mathfrak{c}_{1l} \mathfrak{s}_{Il}}{p \mathfrak{s}_{1l}
\left( \alpha \mathfrak{c}_{1l} + 2 \mathfrak{s}_{1l} \right)} & 0
\\ 0 & 0 & -\frac{ (p-I) \alpha \mathfrak{c}_{Il} - 2 \mathfrak{s}_{Il}}{ \alpha
\mathfrak{c}_{1l} + 2 \mathfrak{s}_{1l}}
\end{pmatrix}
\vspace{0.2cm}\\
\nonumber I&=&5,8,11,\ldots:\\
\nonumber &&(\tilde{M}_{\alpha,I})_l =
\begin{pmatrix}-\frac{ 2 l_1 \left(\mathfrak{c}_{Il} +
\mathfrak{s}_{Il}\right)}{\alpha \mathfrak{c}_{1l} + 2
\mathfrak{s}_{1l}} & -\frac{ 2 l_1 \left(\mathfrak{c}_{Il} +
\mathfrak{s}_{Il}\right)}{\alpha \mathfrak{c}_{1l} + 2
\mathfrak{s}_{1l}} & \frac{ l_1 \mathfrak{s}_{Il}}{\alpha
\mathfrak{c}_{1l} + 2
\mathfrak{s}_{1l}}\\
\frac{ 2 l_2 \left(\mathfrak{c}_{Il} +
\mathfrak{s}_{Il}\right)}{\alpha \mathfrak{c}_{1l} + 2
\mathfrak{s}_{1l}} & \frac{ 2 l_2 \left(\mathfrak{c}_{Il} +
\mathfrak{s}_{Il}\right)}{\alpha \mathfrak{c}_{1l} + 2
\mathfrak{s}_{1l}} & -\frac{ l_2 \mathfrak{s}_{Il}}{\alpha
\mathfrak{c}_{1l} + 2 \mathfrak{s}_{1l}}
\\ \frac{ l_3 (2-l) \alpha p \mathfrak{s}_{Il}}{\alpha
\mathfrak{c}_{1l} + 2 \mathfrak{s}_{1l}} & \frac{ l_3 (2-l) \alpha
p \mathfrak{s}_{Il}}{\alpha \mathfrak{c}_{1l} + 2
\mathfrak{s}_{1l}}  & -\frac{ p \alpha \left(2 \mathfrak{c}_{Il} -
(l_3- \alpha l) \mathfrak{s}_{Il}\right)}{\alpha \mathfrak{c}_{1l}
+ 2 \mathfrak{s}_{1l}}
\end{pmatrix}
\end{eqnarray}
with $l_1=2+l\alpha$, $l_2=2 - (p-l)\alpha$ and $l_3=2+
(p-l)\alpha$. Hence these matrices are well-defined in the limit
of $\alpha \rightarrow 0$ and we can take the limit of
$\tilde{M}_{\alpha,I}$ rather than the whole product in equation
\eqref{SEMES}:
\begin{equation*}
 {(N_{p,I})_j}^k = S_p  \lim_{\alpha\rightarrow 0} \left({E_{p,\alpha}}^{-1} M_{diag,p,\alpha,I}
E_{p,\alpha} \right) S_p\; .
\end{equation*}
We now continue with the right hand side of equation
\eqref{EquiProp} and see that we need to show that
\begin{equation}
\label{KMIKinv}
 K_p M_{p,I} (K_p)^{-1} = \lim_{\alpha\rightarrow 0} \left({E_{p,\alpha}}^{-1} M_{diag,p,\alpha,I}
E_{p,\alpha} \right)\; .
\end{equation}
Therefore we take matrix $K_p$ and $P_p$ from eqns. \eqref{Kexplp} and \eqref{Pexpl}. 
One can simply read off the matrices $M_{p,I}$ from the rows of
$P_p$ (see eq. \eqref{defMp}). We plug these matrices into
the left hand side of equation \eqref{KMIKinv}:
\begin{eqnarray}
\tilde{M}_I &=& K_p M_{p,I} (K_p)^{-1} = (\tilde{M}_I)_0
  \oplus \bigoplus_{l=1}^{p-1} (\tilde{M}_I)_l \\ \vspace{0.5cm}
\nonumber &&(\tilde{M}_I)_0 = (M_{diag,p,\alpha,I})_0  
\end{eqnarray}
\begin{eqnarray}
\nonumber I&=&1,2:\\
\nonumber &&(\tilde{M}_{I})_l =
(-1)^{I(p+l)}\begin{pmatrix} -\frac{ 1 }{\mathfrak{s}_{1l}} &
 -\frac{ 1 }{\mathfrak{s}_{1l}} & 0 \\
\frac{ 1}{\mathfrak{s}_{1l}}  & \frac{ 1 }{\mathfrak{s}_{1l}}  & 0
\\ 0 & 0 & 0
\end{pmatrix} \vspace{0.2cm}\\
\nonumber I&=&3,6,9,\ldots:\\
\nonumber &&(\tilde{M}_I)_l = (-1)^{I}\begin{pmatrix} -\frac{
\left(I  \mathfrak{c}_{Il} + p \mathfrak{s}_{Il} \right)
\mathfrak{s}_{1l} - \mathfrak{c}_{1l} \mathfrak{s}_{Il}}{p
{\mathfrak{s}_{1l}}^2 } & -\frac{I \mathfrak{s}_{1l}
\mathfrak{c}_{Il} - \mathfrak{c}_{1l} \mathfrak{s}_{Il}}{p
{\mathfrak{s}_{1l}}^2 } & 0 \\
\frac{I \mathfrak{s}_{1l} \mathfrak{c}_{Il} - \mathfrak{c}_{1l}
\mathfrak{s}_{Il}}{p {\mathfrak{s}_{1l}}^2 } & \frac{ \left(I
\mathfrak{c}_{Il} - p\mathfrak{s}_{Il}\right)\mathfrak{s}_{1l} -
\mathfrak{c}_{1l} \mathfrak{s}_{Il}}{p {\mathfrak{s}_{1l}}^2 } & 0
\\ 0 & 0 & -\frac{\mathfrak{s}_{Il}}{\mathfrak{s}_{1l}}
\end{pmatrix} \vspace{0.2cm}\\
\nonumber I&=&4,7,10,\ldots:\\
\nonumber &&(\tilde{M}_I)_l = (-1)^{I}\begin{pmatrix} -\frac{
\left((p-I) \mathfrak{c}_{Il} -
p\mathfrak{s}_{Il}\right)\mathfrak{s}_{1l} + \mathfrak{c}_{1l}
\mathfrak{s}_{Il}}{p \mathfrak{s}_{1l}  \mathfrak{s}_{1l} } &
-\frac{ (p-I) \mathfrak{s}_{1l} \mathfrak{c}_{Il} +
\mathfrak{c}_{1l} \mathfrak{s}_{Il}}{p {\mathfrak{s}_{1l}}^2 } & 0 \\
\frac{(p-I) \mathfrak{s}_{1l} \mathfrak{c}_{Il} +
\mathfrak{c}_{1l} \mathfrak{s}_{Il}}{p \mathfrak{s}_{1l}
\mathfrak{s}_{1l}} &
 \frac{\left((p-I)
\mathfrak{c}_{Il} + p\mathfrak{s}_{Il}\right)\mathfrak{s}_{1l} +
\mathfrak{c}_{1l} \mathfrak{s}_{Il}}{p {\mathfrak{s}_{1l}}^2} & 0
\\ 0 & 0 & -\frac{\mathfrak{s}_{Il}}{\mathfrak{s}_{1l}}
\end{pmatrix}
\vspace{0.2cm}\\
\nonumber I&=&5,8,11,\ldots:\\
\nonumber &&(\tilde{M}_I)_l = (-1)^{I}\begin{pmatrix}-\frac{ 2
\left(\mathfrak{c}_{Il} + \mathfrak{s}_{Il}\right)}{
\mathfrak{s}_{1l}} & -\frac{ 2 \left(\mathfrak{c}_{Il} +
\mathfrak{s}_{Il}\right)}{\mathfrak{s}_{1l}} & \frac{
\mathfrak{s}_{Il}}{\mathfrak{s}_{1l}}\\
\frac{ 2 \left(\mathfrak{c}_{Il} + \mathfrak{s}_{Il}\right)}{
\mathfrak{s}_{1l}} & \frac{ 2 \left(\mathfrak{c}_{Il} +
\mathfrak{s}_{Il}\right)}{ \mathfrak{s}_{1l}} & -\frac{
\mathfrak{s}_{Il}}{\mathfrak{s}_{1l}}
\\ 0 & 0  & 0
\end{pmatrix}\; .
\end{eqnarray}
Finally we compare the matrices $\tilde{M}_I$ with the respective
matrices $\tilde{M}_{\alpha,I}$, which constitute the right hand
side of said equation \eqref{KMIKinv}, and notice that the limit
of the latter matrices for $\alpha \rightarrow 0$ yields the
former ones. \qed The
precise connection between $M_{diag,p,\alpha,I}$ and $M_{p,I}$ and
between $K_{diag,p,\alpha}$ and $K_p$ can be clarified a bit more.
We take in eq. \eqref{KMIKinv}
the matrix $K_p$ and its inverse to the other side. As they do not
depend on $\alpha$, we can take them into the limit.
\begin{equation*}
 M_{p,I} = \lim_{\alpha\rightarrow 0} \left((K_p)^{-1} {E_{p,\alpha}}^{-1} M_{diag,p,\alpha,I}
E_{p,\alpha} K_p \right)\; .
\end{equation*}
This gives us already the relation between $M_{diag,p,\alpha,I}$ and
$M_{p,I}$, but we want to have the other one simultaneously, as we
look at the Verlinde formula. \\ $K_{diag,p,\alpha}$ commutes with
$M_{diag,p,\alpha,I}$. So if we insert once the unit matrix, we get
\begin{equation}
\label{MKEKMKEK}
 M_{p,I} = \lim_{\alpha\rightarrow 0} \left((K_p)^{-1} {E_{p,\alpha}}^{-1} K_{diag,p,\alpha} M_{diag,p,\alpha,I}
K_{diag,p,\alpha}^{-1} E_{p,\alpha} K_p \right)\; .
\end{equation}
We then define the matrix
\begin{equation}
\label{DefF}
 F_{p,\alpha}:= K_{diag,p,\alpha}^{-1} E_{p,\alpha} K_p\; .
\end{equation}
This can be easily calculated with equations
\eqref{Kdiagalphadef}, \eqref{Edef} and \eqref{Kexplp}:
\begin{eqnarray*}
 F_{p,\alpha}&=&\boldsymbol{\mathtt{1\!\!l}}_{2\times2}
  \oplus \bigoplus_{j=1}^{p-1} (F_{p,\alpha})_j\\ 
  (F_{p,\alpha})_j&=& \frac{1}{p^3\alpha^2}\begin{pmatrix}
0 & 2 (j\alpha-2)\mathfrak{s}_{1j}\left(\mathfrak{c}_{1j} \alpha +
2 \mathfrak{s}_{1j}\right)&
(-1)^{j+p+1}\sqrt{2p}\left(\mathfrak{c}_{1j} \alpha + 2
  \mathfrak{s}_{1j}\right)\\
0 & 2 ((p-j)\alpha+2)\mathfrak{s}_{1j}\left(\mathfrak{c}_{1j}
\alpha + 2 \mathfrak{s}_{1j}\right)&
(-1)^{j+p}\sqrt{2p}\left(\mathfrak{c}_{1j} \alpha + 2
  \mathfrak{s}_{1j}\right)\\
  1 & - 2 p \alpha \mathfrak{s}_{1j}\left(\mathfrak{c}_{1j}
\alpha + 2 \mathfrak{s}_{1j}\right) & 0
  \end{pmatrix}\; .
\end{eqnarray*}
With these matrices we have the following situation derived from the
limit-Verlinde formula (eq. \eqref{NalphaSMS}):
\begin{eqnarray}
\label{VerlindeVergleich} \mathtt{N}_{p,I}(\alpha)&=&\underbrace{\mathtt{S}_{p,\alpha}E_{p,\alpha}}_{=S_p}
\underbrace{{E_{p,\alpha}}^{-1} K_{diag,p,\alpha}
F_{p,\alpha}}_{=K_p}\cdot \\ \nonumber
&& \cdot {F_{p,\alpha}}^{-1} M_{diag,p,\alpha,I}
F_{p,\alpha} \underbrace{{F_{p,\alpha}}^{-1}
{K_{diag,p,\alpha}}^{-1} E_{p,\alpha}}_{{K_p}^{-1}}
\underbrace{{E_{p,\alpha}}^{-1}
 {\mathtt{S}_{p,\alpha}}^{-1}}_{=S_p}\; .
\end{eqnarray}
One finds that the three matrices in the middle have a regular
limit
\begin{equation}\label{VerlindeVergleich2} \lim_{\alpha\rightarrow 0}\left({F_{p,\alpha}}^{-1}
M_{diag,p,\alpha,I} F_{p,\alpha} \right)=M_{p,I}
\end{equation}
and has the ''generalised'' Verlinde formula for the extended
block diagonalisation method. \\ We have shown in this section,
that both approaches including the indecomposable
representations, which we learned about in the sections
\ref{SecFlohr} and \ref{SecExt}, give the same pre-fusion rules.
Moreover it becomes also clear at this point that Fuchs et al.
found in their work a way to calculate the pre-fusion rules for
irreducible representations in a perhaps mathematically more
appealing and certainly algebraically better motivated way, which
are the same as the ones given by the limit-Verlinde formula.
On the other hand the connection to the work of MF provides
its CFT-side motivation, needs less many different matrices and is
also easier to calculate. Moreover the limit in the
limit-Verlinde formula has now found its justification through
its equality to the ''generalised'' Verlinde formula in our
extension of the block diagonalisation method.

\section{BPZ-Like Closed Forms of the Fusion Rules}
\label{SecBPZ} Finally we want to show the following theorem leading to the pre-fusion algebra of the triplet $\mathcal{W}$-algebra \mbox{$\mathcal{W}(2,(2p-1)^{\otimes 3})$} in a BLZ-like closed form 
\liketheorem{Theorem:}{} The generalised Verlinde formulas in equations \eqref{Nijkalpha} and \eqref{NextSMS} give the following decomposition of the fusion products of irreducible and indecomposable representations:
\begin{eqnarray}
\label{lirrlirr} \left[h_{1,k}\right] \otimes_f \left[h_{1,l}\right] &=& \sum_{m=|k-l|+1\atop \text{step 2}}^{k+l-1} \lambda_m\; ,\\
\label{rirrrirr} \left[h_{1,3p-k}\right] \otimes_f \left[h_{1,3p-l}\right] &=& \sum_{m=|k-l|+1\atop \text{step 2}}^{k+l-1} \lambda_m\; ,\\
\label{lirrrirr} \left[h_{1,k}\right] \otimes_f \left[h_{1,3p-l}\right] &=& \sum_{m=|k-l|+1\atop \text{step 2}}^{k+l-1} \pi_m\; ,\\
\label{indrirr} \left[\widetilde{h_{1,2p-r}}\right] \otimes_f \left[h_{1,2p+s}\right] &=& -\sum_{t=|r-s|+1\atop \text{step 2}}^{\min(r+s-1,\atop 2p-r-s-1)}
 \left[\widetilde{h_{1,2p-t}}\right] + \sum_{{t=\max(1-\left[(r+s) \mod 2 \right],\atop s-r+1)}\atop \text{step 2}}^{\min(p-1+\left[(p+r+s) \mod 2\right],\atop 2p-r-s-1)} \rho_t\; ,\\
\label{ind2p} \left[\widetilde{h_{1,2p-r}}\right] \otimes_f \left[h_{1,2p}\right] &=& \sum_{t=1-\left[r \mod 2 \right]\atop \text{step 2}}^{p-1+\left[(p+r) \mod 2\right]} \rho_t\; ,\\
\label{indp} \left[\widetilde{h_{1,2p-r}}\right] \otimes_f \left[h_{1,p}\right] &=&  \sum_{t=1-\left[r \mod 2 \right]\atop \text{step 2}}^{p-1+\left[(p+r) \mod 2\right]} \rho_{p-t}\; ,\\ 
\label{indlirr} \left[\widetilde{h_{1,2p-r}}\right] \otimes_f \left[h_{1,p-s}\right] &=& \sum_{t=|r-s|+1\atop \text{step 2}}^{\min(r+s-1,\atop 2p-r-s-1)}
 \left[\widetilde{h_{1,p+t}}\right] + \left\{\begin{aligned}\sum_{t=1-\left[(r+s) \mod 2 \right]\atop \text{step 2}}^{r-s-1} \rho_{p-t} &\quad & r>s \\ 0 \quad \quad &\quad & \text{else}\end{aligned}\right\} \\ \nonumber &&
 + \left\{\begin{aligned}\sum_{t=r+s+1\atop \text{step 2}}^{p-1+\left[(p+r+s) \mod 2\right]} \rho_{p-t}&\quad & r+s<p \\ 0 \quad \quad &\quad & \text{else}\end{aligned}\right.\; , \\
\label{indind} \left[\widetilde{h_{1,2p-r}}\right] \otimes_f \left[\widetilde{h_{1,2p-s}}\right] &=& 2  \sum_{t=1-\left[(r+s) \mod 2 \right]\atop \text{step 2}}^{p-1+\left[(p+r+s) \mod 2\right])} \rho_t
\end{eqnarray}
with $0<k,l\leq p$ and $0<r,s<p$ and
\begin{eqnarray}
\label{lambdam} \lambda_m&=&\left\{ \begin{aligned} \left[ h_{1,m} \right] &\quad& 0<m\leq p \\ \left[ h_{1,2p-m} \right] + 2 \left[ h_{1,4p-m} \right] &\quad& p < m < 2p \end{aligned}\right. \; ,\\
\label{pim} \pi_m&=&\left\{ \begin{aligned} \left[ h_{1,3p-m} \right] &\quad& 0<m\leq p \\ \left[ h_{1,m+p} \right] + 2 \left[ h_{1,m-p} \right] &\quad& p < m < 2p \end{aligned}\right. \; , \\
\label{rhot} \rho_t&=&\left\{ \begin{aligned} 2 \left[ h_{1,p} \right] &\quad& t=p \\ 2 \left[ h_{1,2p} \right] &\quad& t=0 \\ 4\left( \left[ h_{1,t} \right]+\left[ h_{1,2p+t} \right] \right) &\quad& 0<t<p
\end{aligned}\right. \; .
\end{eqnarray} \qef
\proof{$\!\!\!\!\!\!$:} The decompositions of the products of irreducible representations in equations \eqref{lirrlirr}-\eqref{lirrrirr}, are proven in section \ref{Projection} to result from the work of Fuchs et al., \protect\cite{Fuchs:2003yu}. There it is shown that the extended version leads to the same decompositions for these products.\\
We bear in mind that the matrices $M_{p,I}$ also fulfil the pre-fusion algebra:
\begin{equation}
\label{MMNM}
 M_{p,I} M_{p,J} = \sum_{K=1}^{3p-1} N_{IJ}^K M_{p,K}\; .
\end{equation}
Because the matrix elements $(M_{p,I})_{1,1}$ are strictly positive, there are no sums of matrices $M_{p,I}$, which are equal to the zero matrix. Thus these decompositions of products $M_{p,I} M_{p,J}$ are unique. \\
For the fusion product in equation \eqref{indrirr} we get the product
\begin{eqnarray*}
 M_{\left[\widetilde{h_{2p-r}}\right]}M_{\left[h_{2p+s}\right]}&=& \mathfrak{P}_0 \oplus \bigoplus_{j=1}^{p-1} \mathfrak{P}_j\\
 \mathfrak{P}_0&=&\begin{pmatrix}
 2p(p-s) &0 \\
 0 & 2(-1)^{r+s}p(p-s)
 \end{pmatrix}\\
 \mathfrak{P}_j&=&\begin{pmatrix}
 0 & (-1)^{r+s+1}\frac{4}{p}\mathfrak{s}_{sj}\left(\mathfrak{c}_{rj}+\mathfrak{s}_{rj}\right) & (-1)^{r+s+p+j+1} \sqrt{\frac{2}{p}}\frac{\mathfrak{s}_{rj}\mathfrak{s}_{sj}}{\mathfrak{s}_{1j}} \\
 0 & 0 & 0 \\
 0 & 0 & 0
 \end{pmatrix}\; .
\end{eqnarray*}
We can now compute the sums of matrices $M_{p,I}$ (see \eqref{MMNM}), which correspond to the proposed fusion rules, and compare them with this product. Plugging in the corresponding $2 \times 2$ blocks of $M_{p,I}$ and $M_{p,I,0}$ (eqns. \eqref{defMp}, \eqref{Pexpl}) for the representations in the decomposition in eq. \eqref{indrirr} directly gives $\mathfrak{P}_0$ in all cases, which need to be distinguished.\\
For the $3 \times 3$ blocks we also start from the proposed decomposition and add up the corresponding blocks, $M_{p,I,j}$. First of all we notice that we encounter only non-zero matrix elements, where also $\mathfrak{P}_j$ is different from zero. This is the case for the indecomposable representations, because here the matrix elements ${(P(p,\alpha)_{s,l})}_{3,1}=0$ for all $0<s,l<p$ (cf. \eqref{Pexpl}). But also for the three possible cases of summands, $\rho_t$, in equation \eqref{rhot} the $3 \times 3$ blocks of the matrices $M_{p,I}$ have or add up to have only the matrix element $(1,2)$ different from zero: The first column of $P(p)_{0,l}$ and  ${(P(p,\alpha)_{s,l})}_{1,1}+{(P(p,\alpha)_{s,l})}_{2,1}$ are zero. The third column of both $P(p)_{0,l}$ and $P(p,\alpha)_{s,l}$ is only non-zero in the rows corresponding to the indecomposable representations. \\
This leaves us for the matrix element $(1,3)$ with the sum of elements $(M_{p,I,j})_{1,3}=(-1)^{p+I+j+1}\sqrt{\frac{2}{p}}\mathfrak{s}_{Ij}$, $I=5,8,11,\ldots$, corresponding to the indecomposable representations appearing in equation \eqref{indrirr}. Here we use identity,
\begin{eqnarray} \label{trigident1}\sum_{I=|r-s|+1 \atop step 2}^{\min(r+s-1 \atop 2p-r-s-1)} \mathfrak{s}_{Ij} &=& \sum_{I=|r-s|+1 \atop step 2}^{r+s-1} \mathfrak{s}_{Ij} = \frac{\mathfrak{s}_{rj}\mathfrak{s}_{sj}}{\mathfrak{s}_{1j}}\; ,
\end{eqnarray}
For the first step index relabeling in part of the sum for the case of $r+s>p$ is needed. Expressed by exponential functions the second step follows from straight forward calculations. We directly get the matrix element $(\mathfrak{P}_j)_{1,3}$.\\
At last we need to get the matrix element $(\mathfrak{P}_j)_{1,2}$, which we will treat in more detail, as it is not so straight forward. From the decomposition in equation \eqref{indrirr} we get for this element:
\begin{equation}
\label{Msum12}\sum_{t=|r-s|+1\atop \text{step 2}}^{\min(r+s-1,\atop 2p-r-s-1)}
 \frac{4}{p} (-1)^t\left(\mathfrak{s}_{tj}+\mathfrak{c}_{tj}\right)\mathfrak{s}_{1j} - \sum_{{t=\max(1-\left[(r+s) \mod 2 \right],\atop s-r+1)}\atop \text{step 2}}^{\min(p-1+\left[(p+r+s) \mod 2\right],\atop 2p-r-s-1)} \frac{4}{p}\mathfrak{s}_{1j} \left\{ \begin{aligned} 1 &\quad& t=0 \\ (-1)^{p+j} &\quad& t=0 \\ 2(-1)^{t}\mathfrak{c}_{tj} &\quad& 0<t<p
\end{aligned}\right. \; . 
\end{equation}
We add the zero,
\begin{eqnarray}
\label{zero1} \frac{4}{p} (-1)^{r+s+1}\mathfrak{s}_{1j} \sum_{t=1-\left[(r+s) \mod 2 \right]\atop \text{step 2}}^{p-1+\left[(p+r+s) \mod 2\right]} \gamma_t &=& 0 \; ,\\
\nonumber \gamma_t=\left\{ \begin{aligned} 1 &\quad& t=0 \\ (-1)^j &\quad& t=p \\  2\mathfrak{c}_{tj} &\quad& \text{else} \end{aligned}\right.\; ,
\end{eqnarray}
and note that $(r+s+1) \mod 2 = t \mod 2$. So the second half of equation \eqref{Msum12} and this zero leaves us with remainders depending on the values of $r$, $s$ and $p$, which we simplify as follows:\\
For the case of $r<s$ we are left with
\begin{equation*}
 \sum_{t=1-\left[(r+s) \mod 2\right]}^{s-r-1} \left\{ \begin{aligned} 1 &\quad& t=0 \\ 2 \mathfrak{c}_{tj} &\quad& \text{else} \end{aligned}\right\} = \sum_{t=r-s+1}^{s-r-1} \mathfrak{c}_{tj}
\end{equation*}
and for $r+s>p$ with
\begin{equation*}
 \sum_{t=2p-r-s+1}^{p-1+\left[(p+r+s) \mod 2\right]} \left\{ \begin{aligned} (-1)^j &\quad& t=p \\ 2 \mathfrak{c}_{tj} &\quad& \text{else} \end{aligned}\right\} = \sum_{t=2p-r-s+1}^{s+r-1} \mathfrak{c}_{tj}\; .
\end{equation*}
The range of the index of these sums then connects directly to the one of the sum in the first half of equation \eqref{Msum12}, which leads in all cases to the following sum of cosines:
\begin{equation*}
\sum_{t=r-s+1 \atop step 2}^{r+s-1} \mathfrak{c}_{tl} = \frac{\mathfrak{s}_{sl}\mathfrak{c}_{rl}}{\mathfrak{s}_{1l}}
\end{equation*}
Finally this results in the matrix element $(\mathfrak{P}_j)_{1,2}$ after applying equation \eqref{trigident1} to the sums of $\mathfrak{s}_{tl}$ once more.
This proofs the decomposition, eq. \eqref{indrirr}.\\
For the decompositions, eqns. \eqref{ind2p} and \eqref{indind}, the $2 \times 2$ blocks are easily checked to be the same on both sides. The $3 \times 3$ blocks in these cases are all on both sides zero. For the decompositions this is seen through equation \eqref{zero1}.\\
The associativity of the pre-fusion product determines the decompositions of the pre-fusion products still left open, eqns. \eqref{indp} and \eqref{indlirr}:
\begin{equation*}
 \left[\widetilde{h_{1,2p-r}}\right] \otimes_f \left[h_{1,p-k}\right] = \left[\widetilde{h_{1,2p-r}}\right] \otimes_f \left[h_{1,2p+k}\right] \otimes_f \left[h_{1,3p-1}\right]
\end{equation*}
with $0 \leq k <p$. The result immediately follows with two pre-fusion products from our previous findings:
\begin{eqnarray*}
\rho_t \otimes_f \left[h_{1,3p-1}\right] &=&\left\{ \begin{aligned} 2 \left[ h_{1,2p} \right] &\quad& t=p \\ 2 \left[ h_{1,p} \right] &\quad& t=0 \\ 4\left( \left[ h_{1,3p-t} \right]+\left[ h_{1,p-t} \right] \right) &\quad& 0<t<p \end{aligned}\right. = \rho_{p-t}\; ,\\
\left[\widetilde{h_{1,2p-t}}\right]\otimes_f \left[h_{1,3p-1}\right] &=&  - \left[\widetilde{h_{1,p+t}}\right] + \rho_{p-t}\; ,
\end{eqnarray*}
\qed
At this point we can now apply the replacement rules from section \ref{SecFlohr} to the pre-fusion rules, eqns. \eqref{lirrlirr}-\eqref{indind}.\\
Only in the fusion products of two irreducible representations the linear combinations needed to be replaced are not immediately visible. We have to distinguish two cases. If $l+m$ in the products, eqns. \eqref{lirrlirr}-\eqref{lirrrirr}, are smaller or equal $p+1$, there are only multiplicities of one appearing in the decomposition, thus no replacements. Otherwise the sum splits into two parts due to the distinction of cases in $\lambda_m$ and $\pi_m$ (eqns. \eqref{lambdam}, \eqref{pim}), as for example for \eqref{lirrlirr}:
\begin{eqnarray*}
\left[h_{1,k}\right] \otimes_f \left[h_{1,l}\right] &=& \sum_{m=|k-l|+1\atop \text{step 2}}^{2p-k-l-1} \left[h_{1,m}\right]+  \underbrace{\sum_{m=2p-k-l+1\atop \text{step 2}}^{p-1+\left[(p+k+l) \mod 2\right]} \left[h_{1,m}\right]}_{=\sum_{m=p+1-\left[(p+k+l) \mod 2\right]\atop \text{step 2}}^{k+l-1} \left[h_{1,2p-m}\right]}\\ &&+  
\sum_{m=p+1+\left[(p+k+l) \mod 2\right]\atop \text{step 2}}^{k+l-1} \biggl(\left[h_{1,2p-m}\right] + 2\left[h_{1,4p-m}\right]\biggr)\\
&=& \sum_{m=|k-l|+1\atop \text{step 2}}^{2p-k-l-1} \left[h_{1,m}\right]+  \underline{\left[h_{1,p}\right]} +  
\sum_{m=p+1+\left[(p+k+l) \mod 2\right]\atop \text{step 2}}^{k+l-1} \left[\widetilde{h_{1,m}}\right]\; ,
\end{eqnarray*} 
where the underlined terms are defined to only appear for odd $k+l+p$. Equation \eqref{lirrrirr} works alike and equation \eqref{rirrrirr} is exactly the same. In summary we get:
\begin{eqnarray}
\label{FusRepla1}
 &&\left[h_{1,k}\right] \otimes_f \left[h_{1,l}\right] = 
 \left[h_{1,3p-k}\right] \otimes_f \left[h_{1,3p-l}\right] \\ \nonumber = &&\left\{ \begin{aligned} \sum_{m=|k-l|+1\atop \text{step 2}}^{k+l-1} \left[h_{1,m}\right] &\quad& 1<k+l\leq p+1 \vspace{0.1in} \\ \underline{\left[h_{1,p}\right]} +  
\sum_{m=|k-l|+1\atop \text{step 2}}^{2p-k-l-1} \left[h_{1,m}\right] + \sum_{m=p+1+\left[(p+k+l) \mod 2\right]\atop \text{step 2}}^{k+l-1} \left[\widetilde{h_{1,m}}\right]&& k+l>p+1\end{aligned} \right.
\end{eqnarray}
\begin{eqnarray*} \left[h_{1,k}\right] \otimes_f \left[h_{1,3p-l}\right] &=& \left\{ \begin{aligned} \sum_{m=|k-l|+1\atop \text{step 2}}^{k+l-1} \left[h_{1,3p-m}\right] &\quad& 1<k+l\leq p+1 \\ \left.  \begin{aligned}&&\underline{\left[h_{1,2p}\right]} +  
\sum_{m=|k-l|+1\atop \text{step 2}}^{2p-k-l-1} \left[h_{1,3p-m}\right] \\ &+&  \sum_{m=p+1+\left[(p+k+l) \mod 2\right]\atop \text{step 2}}^{k+l-1} \left[\widetilde{h_{1,3p-m}}\right]\end{aligned} \right\}&& k+l>p+1\end{aligned} \right.\; .
\end{eqnarray*}
This is in exact correspondence to the fusion rules of irreducible representations of the Virasoro algebra at $c=c_{p,1}$ proposed by Gaberdiel and Kausch in \protect\cite{Gaberdiel:1996kx}. Naturally also the fusion products for $p=2$, $p=3$ and partially for higher $p$, which are calculated in the same paper with their algorithm, are consistent with our result for these fusion products, as well as the following outcome for all other products, after the replacement has been carried out. For the decompositions not involving any indecomposable representations before the replacement (eqns. \eqref{ind2p},\eqref{indp} and \eqref{indind}) we only deal with $\rho_t$, which is replaced by
\begin{equation}
\label{repla1}
 \rho_t \leadsto \left\{ \begin{aligned} 2 \left[ h_{1,p} \right] &\quad& t=p \\ 2 \left[ h_{1,2p} \right] &\quad& t=0 \\ 2 \left[ \widetilde{h_{1,2p-t}} \right] &\quad& 0<t<p
\end{aligned}\right.  \; .
\end{equation}
In the other cases we find
\begin{eqnarray}
\label{FusRepla2a}
\left[\widetilde{h_{1,2p-r}}\right] \otimes_f \left[h_{1,2p+s}\right] &=& \sum_{{t=\max(1-\left[(r+s) \mod 2 \right],\atop s-r+1)}\atop \text{step 2}}^{\min(p-1+\left[(p+r+s) \mod 2\right],\atop 2p-r-s-1)} \tilde{\rho}_t \; ,\\
\label{FusRepla2e} \left[\widetilde{h_{1,2p-r}}\right] \otimes_f \left[h_{1,p-s}\right] &=& \sum_{{t=\max(1-\left[(r+s) \mod 2 \right],\atop s-r+1)}\atop \text{step 2}}^{\min(p-1+\left[(p+r+s) \mod 2\right],\atop 2p-r-s-1)} \hat{\rho}_t
\end{eqnarray}
 with
\begin{eqnarray}
\nonumber
 \tilde{\rho}_t &=&\left\{ \begin{aligned} 2 \left[ h_{1,p} \right] &\quad& t=p \\ 2 \left[ h_{1,2p} \right] &\quad& t=0 \\ \left[ \widetilde{h_{1,2p-t}} \right] &\quad& r-s<t<r+s\\ 2 \left[ \widetilde{h_{1,2p-t}} \right] &\quad& else \; ,
\end{aligned}\right.\\
\nonumber
 \hat{\rho}_t &=&\left\{ \begin{aligned} 2 \left[ h_{1,2p} \right] &\quad& t=p \\ 2 \left[ h_{1,p} \right] &\quad& t=0 \\ \left[ \widetilde{h_{1,p+t}} \right] &\quad& r-s<t<r+s\\ 2 \left[ \widetilde{h_{1,p+t}} \right] &\quad& else \; .
\end{aligned}\right. 
\end{eqnarray}

\section{Conclusion}
In the main part of this paper we have developed an extension of the block diagonalisation method originally introduced by Fuchs et al. leading to a ''generalised'' Verlinde formula. It now additionally includes the fusion products containing indecomposable representations. It reduces to its archetype in every step by simple projection on the components representing the irreducible representations. \\
The block diagonalisation method finds a (linear) algebraic justification through the parallel to the semisimple case. It also performs a simultaneous eigen decomposition of matrices of structure constants of the fusion algebra splitting it into a semisimple algebra and a radical. \\
The S-matrix, $S_p$, for the extended block diagonalisation method is calculated from the parameter dependent S-matrix, $\mathtt{S}_{p,\alpha}$, appearing in the limit-Verlinde formula by a change of basis of chiral vacuum torus amplitudes (cf. eqns. \eqref{Stranschi} and \eqref{SCalphaSCinvalpha}). Thus we have found a CFT-side motivation for this approach because $S_p$ gives the $\mathcal{S}$-transformation of the new basis.\\
We have given a closed form of $\mathtt{S}_{p,\alpha}$ in eq. \eqref{FlohrS}. Although
$\mathtt{S}_{p,\alpha}$ does not diagonalise the fusion
coefficient matrices, it simultaneously diagonalises a set of
matrices depending on $\alpha$ as well. In the limit $\alpha
\rightarrow 0$ these matrices are in accord with the fusion rules, which are known with respect to either the triplet algebra or the Virasoro algebra. \\
We have seen that the pre-fusion rules, which we get from the limit-Verlinde formula or the block diagonalisation method for $c_{p,1}$ models, can not distinguish between indecomposable representations and certain combinations of irreducible representations. As we discussed in section \ref{SecFlohr} in context of the results of the limit-Verlinde formula, this indistinguishability is intrinsic to the
whole calculation on grounds of modular transformations of
characters. At that point it was particularly clear, because the
limit $\alpha \rightarrow 0$ made the forms used for the calculation of $\mathtt{S}_{p,\alpha}$ linearly dependent. \\
Within the in-depth description of this method around the limit-Verlinde formula we have given a detailed formulation of the replacements needed to compensate these intrinsic effects of the linear dependence of characters of the relevant representations. We have collected arguments for these replacements using the quantum numbers of a scaled $\mathfrak{su}(2)$ subalgebra of the triplet $\mathcal{W}$-algebra and unphysical negative coefficients in the decompositions, which are canceled by the replacements.\\
We have shown that the block diagonalisation method is equivalent to the limit-Verlinde formula: It gives the same results starting from the same S-matrix, $\mathtt{S}_{p,\alpha}$. Moreover every matrix in the block diagonalisation method meets its counterpart related to the limit-Verlinde formula, as it is seen in eqns. \eqref{VerlindeVergleich} and \eqref{VerlindeVergleich2}. \\
Through this equivalence the justification of the block diagonalisation method is also true for the limit-Verlinde formula. Here the matrix of simultaneous eigenvectors of the matrices of pre-fusion coefficients is equal to $S_{p,\alpha}K_{diag,p,\alpha}F_{p,\alpha}$ (cf. eq. \eqref{VerlindeVergleich}). Furthermore we immediately get the S-matrix, $\mathtt{S}_{p,\alpha}$, for the limit-Verlinde formula from a ''more canonical'' basis of vacuum torus amplitudes compared to the blockdiagonalisation method. This basis includes the characters of irreducible representations, while the change of basis, which leads to the S-matrix of the block diagonalsiation method, $S_p$, mixes these characters. In addition only the S-matrix, $\mathtt{S}_{p,\alpha}$, is needed in the limit-Verlinde formula to directly calculate the pre-fusion rules.\\
We have calculated the explicit BPZ-like forms for the pre-fusion rules, which result from either of the two methods. As these are not the actual fusion rules for the $c_{p,1}$ models, we have applied the replacement rules to these expressions as well. Hence we finally got our conjecture, what the fusion algebra of these models is, in three different forms: At the end of section \ref{SecFlohr} they are given by the limit-Verlinde formula and subsequently applied replacement rules, which we cast in the form of a case differentiation. With the results of section \ref{SecEqui} we can replace in this conjecture the limit-Verlinde formula by the ''generalised'' Verlinde formula (eq. \eqref{NextSMS}), which gives the same results. Finally we provide explicit expressions, which are given by eqns. \eqref{ind2p}, \eqref{indp} and \eqref{indind} with replacement \eqref{repla1}, eq. \eqref{FusRepla1} and eqns. \eqref{FusRepla2a}-\eqref{FusRepla2e}.\\
There are many parallels of the $c_{p,1}$ models to rational conformal field theories with completely reducible symmetry algebras and thus with semisimple fusion algebras. Most importantly here a S-matrix can be calculated from a basis of chiral vacuum torus amplitudes and used in a generalisation of the Verlinde formula. It provides us with a well-founded proposition for the fusion rules of the $c_{p,1}$ series, justified by many indications, and gives the -- seemingly for higher $p$ -- correct result. \\
There are several important questions, which our work also approaches.\\ 
For which logarithmic conformal field theories do we expect to find a generalisation of the Verlinde formula? How does it look like? Everything points to the need of a basis of chiral vacuum torus amplitudes including the characters of irreducible representations. The $\mathcal{S}$-transformation of this basis then determines the S-matrix. Probably it would again depend on a parameter and for the limit of this parameter to zero become the set of characters of all irreducible and indecomposable representations.\\
Then, of course, the question of a proof of the presented Verlinde-like formulas and the fusion rules, which follow from them after certain well-defined replacements, suggests itself and is connected to the previous questions. We have seen many similarities to semisimple fusion algebras. This suggests things to be not so different with or without semisimplicity. Still there is the obstacle of the needed replacements, which need to be better understood.\\
The Verlinde formula in rational conformal field theories has found deep roots in the algebraic geometric background of these theories. The work on it has gone far into their rigorous formulations. Our findings here give confidence, that this rigorousness is also what lies ahead of us for certain logarithmic conformal field theories.

\vspace{.2in}
\paragraph{\textbf{Acknowledgements:}} We would like to thank Hendrik Adorf for reading the script of this paper with great care. The work of MF is partially supported by the European Union network HPRN-CT-2002-00325 (EUCLID). 

\appendix
\section{Details on the Choice Inherent in the Forms ${\tilde{\chi}}_{\lambda,p}(\alpha)$}\label{appchoice}
In this section details are given on the considerations leading to the linear independent set of forms representing irreducible and indecomposable representations, which is used in section \ref{SecFlohr} to calculate the S-matrix $\mathtt{S}_{p,\alpha}$. \\
In this context we have to recall a essential part of the calculation of the partition function of the $c_{p,1}$ models in \protect\cite{Flo96} from the characters of the representations of the triplet algebra: Further forms have been introduced there to solve the problems arising from the specific  modular transformation properties of these characters. In fact the characters of the indecomposable representations are split into a sum:
\begin{eqnarray}\label{splitchar}
{\chi^{\mathcal{R}}}_{\lambda,p}&=&\frac{2}{p}\left[ (p-\lambda)\chi^{\mathcal{R}+}_{\lambda,p}(\alpha)+\lambda \chi^{\mathcal{R}-}_{\lambda,p}(\alpha)\right]\; ,\\
 \nonumber
\chi^{\mathcal{R}+}_{\lambda,p}(\alpha)&=&\frac{1}{\eta}\left[
\Theta_{\lambda,p}+\mathrm{i} \alpha \lambda (\nabla
\Theta)_{\lambda,p}\right]\; , \\
\nonumber
\chi^{\mathcal{R}-}_{\lambda,p}(\alpha)&=&\frac{1}{\eta}\left[
\Theta_{\lambda,p}-\mathrm{i} \alpha (p-\lambda) (\nabla
\Theta)_{\lambda,p}\right]\; ,
\end{eqnarray}
where $(\nabla \Theta)_{\lambda,p}$ is
\begin{equation}\label{NablaTheta}
(\nabla \Theta)_{\lambda,p} = \mathrm{i} \tau (\partial \Theta)_{\lambda,k}
= \frac{1}{2 \pi} \log(q)(\partial \Theta)_{\lambda,k}\; .
\end{equation}
The partition function is given in terms of the characters of irreducible representations and these forms and stays modular invariant for $\alpha \rightarrow 0$.\\
Concerning our goal to find a $3p-1 \times 3p-1$ S-matrix for the $c_{p,1}$ models there are now linearly independent sets of characters of irreducible representations and linear combination of
$\chi^{\mathcal{R}+}_{\lambda,p}(\alpha)$ and
$\chi^{\mathcal{R}-}_{\lambda,p}(\alpha)$ with $(3p-1)$ elements. What is more, these sets close under modular transformations of
their argument, i.e. any of these $3p-1$ forms evaluated at
$\gamma\tau$, with $\gamma \in SL(2,\mathbb{Z})$, can be written
as a linear combination of the same forms evaluated at $\tau$.\\
The possible $p-1$ linear combinations of $\chi^{\mathcal{R}+}_{\lambda,p}(\alpha)$ and
$\chi^{\mathcal{R}-}_{\lambda,p}(\alpha)$ are parametrised by $x\in \mathbb{C}$ in the following way:
\begin{equation}\label{lincombchar}
{\tilde{\chi}}_{\lambda,p}(\alpha,x)=\frac{2}{p}\left[
(p+x-\lambda)\chi^{\mathcal{R}+}_{\lambda,p}(\alpha)+(\lambda-x)\chi^{\mathcal{R}-}_{\lambda,p}(\alpha)\right]\; .
\end{equation}
The results for the fusion rules do not depend on the choice of $x$. The forms
${\tilde{\chi}}_{\lambda,p}(\alpha,x)$ surely depends on $x$.
However, when we insert the forms
$\chi^{\mathcal{R}+}_{\lambda,p}(\alpha)$ and
$\chi^{\mathcal{R}-}_{\lambda,p}(\alpha)$ (eq. \eqref{splitchar})
into equation \eqref{lincombchar}, it emerges, that it only
depends on the product of $x$ and $\alpha$:
\begin{equation*}
{\tilde{\chi}}_{\lambda,p}(\alpha,x)=\frac{1}{\eta}\left[
2\Theta_{\lambda,p} + 2 x \mathrm{i} \alpha (\nabla
\Theta)_{\lambda,p}\right]\; .
\end{equation*}
We can redefine $\alpha$ in a convenient way to incorporate $x$.
Because we take the limit $\alpha \rightarrow 0$ at the end, this
does not change the results.\\
For the following $x=-\mathrm{i}/2$ is chosen, which corresponds for $p=2$ to the
choice made in \protect\cite{Flohr:1996vc}:
\begin{equation}
\label{ChiTildeDef}
{\tilde{\chi}}_{\lambda,p}(\alpha)={\tilde{\chi}}_{\lambda,p}(\alpha,-\mathrm{i})=\frac{1}{\eta}\left[
2\Theta_{\lambda,p} + \alpha (\nabla \Theta)_{\lambda,p}\right]\; .
\end{equation}
The factor $2/p$ appears in equation \eqref{lincombchar} in
contrast to \protect\cite{Flohr:1996vc} in order to have a multiplicity  - which a priori may be chosen - of 2 in front of the
$\Theta_{\lambda,p}/\eta$ term in
${\tilde{\chi}}_{\lambda,p}(\alpha,-\mathrm{i})$, instead of a multiplicity
of $p$. The result at the end depends on the choice of the
multiplicity. Another multiplicity in
${\tilde{\chi}}_{\lambda,p}(\alpha,x)$ leads qualitatively to the
correct fusion rules, but with different multiplicities. Our
choice is the one, for which the forms
${\tilde{\chi}}_{\lambda,p}(\alpha)$ become the characters of the
indecomposable representations for $\alpha \rightarrow 0$.

\section{Fusion rules for $p=2$ and $p=3$}
\label{appfusp2p3}
In tables \ref{prefusp2} and \ref{prefusp3} we give the pre-fusion rules resulting from the limit-Verlinde formula -- and thus also from the extended block diagonalisation method -- for the $c_{2,1}=-2$ and the $c_{3,1}=-7$ models, respectively. The fusion rules, which are the outcome, after the replacement rules around eq. \eqref{replincomb} have been applied, are listed in tables \ref{fusp2} and \ref{fusp3}.\\
In the next subsection we go through a few examples of the application of these replacement rules for $p=3$.  
\begin{table}
	\centering
		\begin{tabular}{l|ccccc}
	$\otimes_f$	& $\left[-\frac{1}{8}\right]$ & $\left[\frac{3}{8}\right]$ & $\left[0\right]$ & $\left[1\right]$ & $\left[\tilde{0}\right]$ \\ \hline

$\left[-\frac{1}{8}\right]$ & $2\left[0\right]+2\left[1\right]$ & $2\left[0\right]+2\left[1\right]$ & $\left[-\frac{1}{8}\right]$ & $\left[\frac{3}{8}\right]$ & $2\left[-\frac{1}{8}\right]+2\left[\frac{3}{8}\right]$\\

$\left[\frac{3}{8}\right]$ & $2\left[0\right]+2\left[1\right]$ & $2\left[0\right]+2\left[1\right]$ & $\left[\frac{3}{8}\right]$ & $\left[-\frac{1}{8}\right]$ & $2\left[-\frac{1}{8}\right]+2\left[\frac{3}{8}\right]$\\

$\left[0\right]$ & $\left[-\frac{1}{8}\right]$ & $\left[\frac{3}{8}\right]$ & $\left[0\right]$ & $\left[1\right]$ & $\left[\tilde{0}\right]$\\

$\left[1\right]$ & $\left[\frac{3}{8}\right]$ & $\left[-\frac{1}{8}\right]$ & $\left[1\right]$ & $\left[0\right]$ & $4\left[0\right]+4\left[1\right]-\left[\tilde{0}\right]$\\
$\left[\tilde{0}\right]$ & $2\left[-\frac{1}{8}\right]+2\left[\frac{3}{8}\right]$ & $2\left[-\frac{1}{8}\right]+2\left[\frac{3}{8}\right]$ & $\left[\tilde{0}\right]$ &  $4\left[0\right]+4\left[1\right]-\left[\tilde{0}\right]$ & $8\left[0\right]+8\left[1\right]$
\end{tabular}
	\caption{Pre-fusion rules for p=2}
	\label{prefusp2}
\end{table}
\begin{table}
	\centering
		\begin{tabular}{l|ccccc}
		$\otimes_f$	& $\left[-\frac{1}{8}\right]$ & $\left[\frac{3}{8}\right]$ & $\left[0\right]$ & $\left[1\right]$ & $\left[\tilde{0}\right]$ \\ \hline

$\left[-\frac{1}{8}\right]$ & $\left[\tilde{0}\right]$ & $\left[\tilde{0}\right]$ & $\left[-\frac{1}{8}\right]$ & $\left[\frac{3}{8}\right]$ & $2\left[-\frac{1}{8}\right]+2\left[\frac{3}{8}\right]$\\

$\left[\frac{3}{8}\right]$ & $\left[\tilde{0}\right]$ & $\left[\tilde{0}\right]$ & $\left[\frac{3}{8}\right]$ & $\left[-\frac{1}{8}\right]$ & $2\left[-\frac{1}{8}\right]+2\left[\frac{3}{8}\right]$\\

$\left[0\right]$ & $\left[-\frac{1}{8}\right]$ & $\left[\frac{3}{8}\right]$ & $\left[0\right]$ & $\left[1\right]$ & $\left[\tilde{0}\right]$\\

$\left[1\right]$ & $\left[\frac{3}{8}\right]$ & $\left[-\frac{1}{8}\right]$ & $\left[1\right]$ & $\left[0\right]$ & $\left[\tilde{0}\right]$\\
$\left[\tilde{0}\right]$ & $2\left[-\frac{1}{8}\right]+2\left[\frac{3}{8}\right]$ & $2\left[-\frac{1}{8}\right]+2\left[\frac{3}{8}\right]$ & $\left[\tilde{0}\right]$ &  $\left[\tilde{0}\right]$ & $4\left[\tilde{0}\right]$
\end{tabular}
	\caption{Fusion rules for p=2}
	\label{fusp2}
\end{table}

\begin{table}
\centering
\begin{tabular}{l|cccc}
$\otimes_f$ & $\left[-\frac{1}{3}\right]$ & $\left[\frac{5}{12}\right]$ & $\left[0\right]$ & $\left[1\right]$ \\
 \hline 
$\left[-\frac{1}{3}\right]$ & $\left[-\frac{1}{3}\right]+2\left[0\right]+2\left[1\right]$ 
\\ 
$\left[\frac{5}{12}\right]$ & $\left[\frac{5}{12}\right]+ 2\left[-\frac{1}{4}\right]+ 2\left[\frac{7}{4}\right]$ & $\left[-\frac{1}{3}\right]+2\left[0\right]+2\left[1\right]$ 
\\ 
$\left[0\right]$ & $\left[-\frac{1}{3}\right]$ & $\left[\frac{5}{12}\right]$ & $\left[0\right]$ 
\\
$\left[1\right]$ & $2\left[0\right]+2\left[1\right]$  & $2\left[-\frac{1}{4}\right]+ 2\left[\frac{7}{4}\right]$ & $\left[1\right]$ & $\left[0\right]+\left[-\frac{1}{3}\right]$ \\
$\left[\tilde{0}\right]$ & $2\left[-\frac{1}{3}\right]+4\left[0\right]+4\left[1\right]$ & $2\left[\frac{5}{12}\right]+ 4\left[-\frac{1}{4}\right]+ 4\left[\frac{7}{4}\right]$ & $\left[\tilde{0}\right]$ &  $2\left[-\frac{1}{3}\right] +4\left[0\right] \atop +4\left[1\right] -\left[\tilde{0}\right]$ \\
$\left[-\frac{1}{4}\right]$ & $2\left[-\frac{1}{4}\right]+ 2\left[\frac{7}{4}\right]$ & $2\left[0\right]+2\left[1\right]$ & $\left[-\frac{1}{4}\right]$ & $\left[\frac{5}{12}\right] +\left[\frac{7}{4}\right]$\\
$\left[\frac{7}{4}\right]$ & $\left[\frac{5}{12}\right]$ & $\left[-\frac{1}{3}\right]$ & $\left[\frac{7}{4}\right]$ & $\left[-\frac{1}{4}\right]$\\
\vspace{0.2in} $\left[\widetilde{-\frac{1}{4}}\right]$ & $2\left[\frac{5}{12}\right]+ 4\left[-\frac{1}{4}\right] +4\left[\frac{7}{4}\right]$ & $2\left[-\frac{1}{3}\right]+4\left[0\right]+4\left[1\right]$ & $\left[\widetilde{-\frac{1}{4}}\right]$ & $2\left[\frac{5}{12}\right] + 4\left[-\frac{1}{4}\right]\atop +4\left[\frac{7}{4}\right] -\left[\widetilde{-\frac{1}{4}}\right]$
\end{tabular}
\begin{tabular}{l|cccc}
		$\otimes_f$	& $\left[\tilde{0}\right]$ & $\left[-\frac{1}{4}\right]$ & $\left[\frac{7}{4}\right]$ & $\left[\widetilde{-\frac{1}{4}}\right]$ \\ \hline  
$\left[\tilde{0}\right]$ & $4\left[-\frac{1}{3}\right]+8\left[0\right]+8\left[1\right]$ \\
$\left[-\frac{1}{4}\right]$ & $2\left[\frac{5}{12}\right] +\left[\widetilde{-\frac{1}{4}}\right]$ & $\left[-\frac{1}{3}\right] +\left[0\right]$ \\
$\left[\frac{7}{4}\right]$ & $4\left[-\frac{1}{4}\right]+4\left[\frac{7}{4}\right] -\left[\widetilde{-\frac{1}{4}}\right]$ & $\left[1\right]$ & $\left[0\right]$ \\
$\left[\widetilde{-\frac{1}{4}}\right]$ & $4\left[\frac{5}{12}\right] + 8\left[-\frac{1}{4}\right]+8\left[\frac{7}{4}\right]$ & $2\left[-\frac{1}{3}\right] + \left[\tilde{0}\right]$ & $4\left[0\right] +4\left[1\right] -\left[\tilde{0}\right]$ & $4\left[-\frac{1}{3}\right] + 8\left[0\right] +8\left[1\right]$
\end{tabular}
	\caption{Pre-fusion rules for p=3}
	\label{prefusp3}
\end{table}

\begin{table}
	\centering
		\begin{tabular}{l|ccccc}
$\otimes_f$ & $\left[-\frac{1}{3}\right]$ & $\left[\frac{5}{12}\right]$ & $\left[0\right]$ & $\left[1\right]$ \\ \hline 
$\left[-\frac{1}{3}\right]$ & $\left[-\frac{1}{3}\right]+\left[\tilde{0}\right]$ 
\\ 
$\left[\frac{5}{12}\right]$ & $\left[\frac{5}{12}\right]+ \left[\widetilde{-\frac{1}{4}}\right]$ & $\left[-\frac{1}{3}\right]+\left[\tilde{0}\right]$ 
\\ 
$\left[0\right]$ & $\left[-\frac{1}{3}\right]$ & $\left[\frac{5}{12}\right]$ & $\left[0\right]$ 
\\
$\left[1\right]$ & $\left[\tilde{0}\right]$  & $\left[\widetilde{-\frac{1}{4}}\right]$ & $\left[1\right]$ & $\left[0\right]+\left[-\frac{1}{3}\right]$ \\
$\left[\tilde{0}\right]$ & $2\left[-\frac{1}{3}\right]+2\left[\tilde{0}\right]$ & $2\left[\frac{5}{12}\right]+ 2\left[\widetilde{-\frac{1}{4}}\right]$ & $\left[\tilde{0}\right]$ &  $2\left[-\frac{1}{3}\right] + \left[\tilde{0}\right]$ \\
$\left[-\frac{1}{4}\right]$ & $\left[\widetilde{-\frac{1}{4}}\right]$ & $\left[\tilde{0}\right]$ & $\left[-\frac{1}{4}\right]$ & $\left[\frac{5}{12}\right] +\left[\frac{7}{4}\right]$\\
$\left[\frac{7}{4}\right]$ & $\left[\frac{5}{12}\right]$ & $\left[-\frac{1}{3}\right]$ & $\left[\frac{7}{4}\right]$ & $\left[-\frac{1}{4}\right]$\\
\vspace{0.2in} $\left[\widetilde{-\frac{1}{4}}\right]$ & $2\left[\frac{5}{12}\right]+ 2\left[\widetilde{-\frac{1}{4}}\right]$ & $2\left[-\frac{1}{3}\right]+2\left[\tilde{0}\right]$ & $\left[\widetilde{-\frac{1}{4}}\right]$ & $2\left[\frac{5}{12}\right] + \left[\widetilde{-\frac{1}{4}}\right]$ \\
$\otimes_f$	& $\left[\tilde{0}\right]$ & $\left[-\frac{1}{4}\right]$ & $\left[\frac{7}{4}\right]$ & $\left[\widetilde{-\frac{1}{4}}\right]$ \\\hline  
$\left[\tilde{0}\right]$ & $4\left[-\frac{1}{3}\right]+4\left[\tilde{0}\right]$ \\
$\left[-\frac{1}{4}\right]$ & $2\left[\frac{5}{12}\right] +\left[\widetilde{-\frac{1}{4}}\right]$ & $\left[-\frac{1}{3}\right] +\left[0\right]$ \\
$\left[\frac{7}{4}\right]$ & $\left[\widetilde{-\frac{1}{4}}\right]$ & $\left[1\right]$ & $\left[0\right]$ \\
$\left[\widetilde{-\frac{1}{4}}\right]$ & $4\left[\frac{5}{12}\right] + 4\left[\widetilde{-\frac{1}{4}}\right]$ & $2\left[-\frac{1}{3}\right] + \left[\tilde{0}\right]$ & $\left[\tilde{0}\right]$ & $4\left[-\frac{1}{3}\right] + 4\left[\tilde{0}\right]$
\end{tabular}
	\caption{Fusion rules for p=3}
	\label{fusp3}
\end{table}

\subsection{Demonstration of Replacement Rules for $p=3$}
\label{appDemRRp3}
The matrix $\mathtt{S}_{3,\alpha}$ reads
\begin{equation}
\label{exampSalphap3}
\begin{pmatrix}
 \frac{1}{2}\hat{r} &\frac{1}{2}\hat{r} &\hat{r} & \hat{r}& 0 & \hat{r} & \hat{r} & 0\\
 \frac{1}{2}\hat{r} &-\frac{1}{2}\hat{r} &\hat{r} & \hat{r}& 0 & -\hat{r} & -\hat{r} & 0\\
\frac{1}{6}\hat{r} & \frac{1}{6}\hat{r} &
-\frac{1}{6}\hat{r}-\hat{s} & -\frac{1}{6}\hat{r}-\hat{s} &
\frac{1}{2}\hat{s} & -\frac{1}{6}\hat{r}+\hat{s} &
-\frac{1}{6}\hat{r}+\hat{s} & -\frac{1}{2}\hat{s}\\
\frac{1}{3}\hat{r} & \frac{1}{3}\hat{r} &
-\frac{1}{3}\hat{r}+\hat{s} & -\frac{1}{3}\hat{r}+\hat{s} &
-\frac{1}{2}\hat{s} & -\frac{1}{3}\hat{r}-\hat{s} &
-\frac{1}{3}\hat{r}-\hat{s} & \frac{1}{2}\hat{s}\\
\hat{r} & \hat{r} & -\hat{r}+\hat{t} &
-\hat{r}-\frac{1}{2}\hat{t}
& 0 & -\hat{r}-\frac{1}{2}\hat{t} & -\hat{r}+\hat{t} & 0 \\
\frac{1}{3}\hat{r} & -\frac{1}{3}\hat{r} &
-\frac{1}{3}\hat{r}+\hat{s} & -\frac{1}{3}\hat{r}+\hat{s} &
-\frac{1}{2}\hat{s} & \frac{1}{3}\hat{r}+\hat{s} &
\frac{1}{3}\hat{r}+\hat{s} & -\frac{1}{2}\hat{s}\\
\frac{1}{6}\hat{r} & -\frac{1}{6}\hat{r} &
-\frac{1}{6}\hat{r}-\hat{s} & -\frac{1}{6}\hat{r}-\hat{s} &
\frac{1}{2}\hat{s} & \frac{1}{6}\hat{r}-\hat{s} &
\frac{1}{6}\hat{r}-\hat{s} & \frac{1}{2}\hat{s}\\ 
\hat{r} & -\hat{r} & -\hat{r}-\hat{t} &
-\hat{r}+\frac{1}{2}\hat{t} & 0 & \hat{r}-\frac{1}{2}\hat{t} &
\hat{r}+\hat{t} & 0
\end{pmatrix}
\end{equation}
with $\hat{r}=\sqrt{6}/3$, $\hat{s}=\sqrt{2}/(3\alpha)$ and
$\hat{t}=\alpha\sqrt{2}$. Eq. \eqref{Nijkalpha} then gives
pre-fusion rules listed in table \ref{prefusp3}.\\ It is
worth going through some particular fusion products to see the problems arising through the
ambiguities in the limit $\alpha \rightarrow 0$. In this example
there are two indecomposable representations and two corresponding
identities of their characters:
\begin{equation*}
2\chi^+_{i,3}+2\chi^-_{i,3}=\chi^{\mathcal{R}}_{i,3} \quad i=1,2\;
.
\end{equation*}
These ''translate'' to identities of representations, which shall
symbolise their indistinguishabileness in this calculation:
\begin{eqnarray}
\nonumber
2\left[-\frac{1}{4}\right]+2\left[\frac{7}{4}\right]&=&\left[\widetilde{-\frac{1}{4}}\right]
\; ,\\ \label{p3lincomb}
2\left[0\right]+2\left[1\right]&=&\left[\tilde{0}\right]\; .
\end{eqnarray}
Quite typical is the following product:
\begin{equation*}
\left[-\frac{1}{4}\right] \otimes^f \left[-\frac{1}{3}\right] = 2
\left[-\frac{1}{4}\right] + 2 \left[\frac{7}{4}\right]\; .
\end{equation*}
Here the first identity in eq. \eqref{p3lincomb} is used to get
the desired result $\left[\widetilde{-\frac{1}{4}}\right]$.
Replacements of this kind are still quite comprehensible. But
there are several results for other fusion products like
\begin{equation}
\label{p3fusex1} \left[\tilde{0}\right] \otimes^f \left[1\right] =
4 \left[0\right] - \left[\tilde{0}\right] + 2
\left[-\frac{1}{3}\right] + 4 \left[1\right]\; ,
\end{equation}
which catch one's eye because of a disturbing minus sign. But it
also contains the latter of the linear combinations in eq.
\eqref{p3lincomb} in a sufficiently high multiplicity, so that we
can mend this problem by a calculation on the level of characters.
Equation \eqref{p3fusex1} then yields
\begin{equation*}
 2 \left[\tilde{0}\right] - \left[\tilde{0}\right] + 2 \left[-\frac{1}{3}\right] = \left[\tilde{0}\right] + 2
 \left[-\frac{1}{3}\right]\; .
\end{equation*}
This kind of calculation must be done in several fusion products
given in table \ref{prefusp3}. For those products one finally gets 
the fusion rules for $\mathcal{W}$-algebra representations, which
are listed in table \ref{fusp3} and are consistent with the fusion
rules calculated for the Virasoro modules in
\protect\cite{Gaberdiel:1996kx}.

\section{Findings for the case $p=2$: $S_2$, $C_2(\alpha)$ and $K_2$} \label{SCKfromp2}
Here we look at the simplest case, $p=2$, and search the matrix
$C_2(\alpha)$, for which we will have
\begin{equation}
\label{C2def}
 S_2 = C_2(\alpha) \mathtt{S}_{2,\alpha}
 {C_2}^{-1}(\alpha)\; ,
\end{equation}
where the matrix $\mathsf{S}(2)$ (eq. \eqref{SirrDef}) appears as a block in $S_2$. The fifth line of $S_2$ is yet undetermined as well. We write
\begin{equation}
\label{S2form}
  S_2=\begin{pmatrix}
     \mathsf{S}(2)& \begin{matrix}
     0 \\
     0 \\
     0 \\
     0 \\
                  \end{matrix} \\
\begin{matrix} s_1 & s_2 & s_2 & s_4 \end{matrix} & s_5
               \end{pmatrix}
\end{equation}
with $s_5 \neq 0$. Notice that $S_2$ -- even if known completely -- leaves several possible $C_2(\alpha)$ fulfilling equation
\eqref{C2def} and a couple of conditions we want to impose on $C_2(\alpha)$ such as block diagonality.\\
Just looking at $p=2$ these matrices will seem equally suitable. We will only be able to single out a specific
$C_2(\alpha)$, when we ask, for which we can find a generalisation
to arbitrary $p$. \\
A glance on the
eigenvalues of the matrices $S_{2}$ and $\mathtt{S}_{2,\alpha}$, will give first restrictions on the matrix $S_{2}$.
\begin{table}
  \begin{tabular}{lcccccc}
  \multicolumn{3}{c}{$1$}&$\quad$& $-1$& \\ \hline
$\begin{pmatrix}
                       1 \\ 1 \\ 0 \\ 0
                         \end{pmatrix}$&$\begin{pmatrix}
                       2 \\ 0 \\ 1 \\ 0
                         \end{pmatrix}$&$\begin{pmatrix}
                       2 \\ 0 \\ 0 \\ 1
                         \end{pmatrix}$&$\quad$&$\begin{pmatrix}
                       -2 \\ 2 \\ 1 \\ 1
                         \end{pmatrix}$&
\end{tabular}
\caption{Eigenvalues and eigenvectors of $\mathsf{S}(2)$}
\label{appeigvec1}
\end{table}
\begin{table}
  \begin{tabular}{cccccc}
\multicolumn{3}{c}{$1$}&$\quad$&\multicolumn{2}{c}{$-1$}
\\  \hline  $\begin{pmatrix}
                       2 \\ 0 \\ 1 \\ 0 \\ 2-\alpha
                         \end{pmatrix}$&$\begin{pmatrix}
                       2 \\ 0 \\ 0 \\ 1 \\ 2+\alpha
                         \end{pmatrix}$&$\begin{pmatrix}
                       1 \\ 1 \\ 0 \\ 0 \\ 0
                         \end{pmatrix}$&$\quad$&$\begin{pmatrix}
                       -1 \\ 1 \\ 1 \\ 0 \\ 2+\alpha
                         \end{pmatrix}$&$\begin{pmatrix}
                       -1 \\ 1 \\ 0 \\ 1 \\ 2-\alpha
                         \end{pmatrix}$
\end{tabular}
\caption{Eigenvalues and eigenvectors of $\mathtt{S}_{2,\alpha}$}
\label{appeigvec2}
\end{table}
\begin{table}
  \begin{tabular}{ccccccc}
\multicolumn{3}{c}{$1$}&$\quad$& $-1$ &$\quad$& $s_5$
\\ \hline $\begin{pmatrix}
                       \frac{2s_2-s_3}{s_1+s_2} \\  -\frac{2s_1+s_3}{s_1+s_2} \\ 1 \\ 0 \\ 0
                         \end{pmatrix}$&$\begin{pmatrix}
                       \frac{2s_2-s_4}{s_1+s_2} \\ -\frac{2s_1+s_3}{s_1+s_2} \\ 0 \\ 1 \\ 0
                         \end{pmatrix}$&$\begin{pmatrix}
                       \frac{1-s_5}{s_1+s_2} \\ \frac{1-s_5}{s_1+s_2} \\ 0 \\ 0 \\ 1
                         \end{pmatrix}$&$\quad$&$\begin{pmatrix}
                       -2 \\ 2 \\ 1 \\ 1 \\ \frac{-2s_2-s_3-s_4+2s_1}{s_5+1}
                         \end{pmatrix}$&$\quad$&$\begin{pmatrix}
                       0 \\ 0 \\ 0 \\ 0 \\ 1
                         \end{pmatrix}$
\end{tabular}
\caption{Eigenvalues and eigenvectors of $S_2$ for $s_5 \neq -1$}
\label{appeigvec3}
\end{table}
\begin{table}
\begin{tabular}{cccccc}
\multicolumn{3}{c}{$1$}&$\quad$& \multicolumn{2}{c}{$-1$}
\\ \hline $\begin{pmatrix}
                       1 \\ 1 \\ 0 \\ 0 \\  \frac{s_4}{4} + \frac{s_3}{4} + s_2
                         \end{pmatrix}$&$\begin{pmatrix}
                       2 \\ 0 \\ 1 \\ 0 \\ \frac{s_4}{2} + s_3 + s_2
                         \end{pmatrix}$&$\begin{pmatrix}
                       2 \\ 0 \\ 0 \\ 1 \\ s_4 + \frac{s_3}{2} + s_2
                         \end{pmatrix}$&$\quad$& $\begin{pmatrix}
                       -2 \\ 2 \\ 1 \\ 1 \\ 0
                         \end{pmatrix}$&$\begin{pmatrix}
                       0 \\ 0 \\ 0 \\ 0 \\ 1
                         \end{pmatrix}$
\end{tabular}
\caption{Eigenvalues and eigenvectors of $S_2 \text{ for } s_5 = -1$ and $s_1 = \frac{2s_2 + s_3 + s_4}{2}$}
\label{appeigvec4}
\end{table}
The eigenvalues and eigenvectors of the different S-matrices for
$p=2$ are listed in tables \ref{appeigvec1}-\ref{appeigvec4}. \\
$S_{2}$ and $\mathtt{S}_{2,\alpha}$ are both diagonalisable. The former one
has a three dimensional eigenspaces for the eigenvalue $1$ and two
one dimensional eigenspace for the eigenvalue $-1$ and for the
eigenvalue $s_5$, respectively. For the latter one it is not so different. It
has eigenvalues $1$ and $-1$ belonging to eigenspaces with
dimensions three and two, respectively. \\
If now $S_2$ is chosen, so that the fifth eigenvalue -- and matrix
element -- $s_5$ is also $-1$, the matrices $S_2$ and
$\mathtt{S}_{2,\alpha}$ are diagonalised to the same matrix
\begin{equation}
\label{DSDef} D_S= \begin{pmatrix}
  1 &0 &0 &0 &0 \\
  0 &1 &0 &0 &0 \\
  0 &0 &1 &0 &0 \\
  0 &0 &0 &-1 &0 \\
  0 &0 &0 &0 &-1 \\
 \end{pmatrix} \; .
\end{equation}
But this is only the case, if we also have $s_1 =
s_2 + s_3/2 +s_4/2$. Otherwise we get an undiagonalisable matrix. This is also apparent from fifth component of the forth eigenvector of $S_2$ for $s_5 \neq -1$ listed in table \ref{appeigvec3}, which is not defined then.\\
With this condition and $s_5=-1$ the
eigenvectors have the same first four components as the
eigenvectors of the smaller matrix $\mathsf{S}(2)$ (compare tables \ref{appeigvec1} and \ref{appeigvec4}). \\
To continue to determine $S_2$ we recall the block diagonal form of $M_{p,I}$ in equation \eqref{defMp} and that we have taken $K_p$ to be block diagonal in our argumentation in section \ref{SecConSP}, where we fixed $K_p$ at the end taking the simplest choice. So $S_2$
should already block diagonalise the fusion rules. For this we take the results in table \ref{prefusp2}, which have been calculated following section \ref{SecFlohr}. Here we only need the first matrix of fusion coefficients giving the fusion rules for product with $\left[ - 1/8\right]$:
\begin{equation}
N_{2,1} =
\begin{pmatrix} 
0 & 0 & 2 & 2 & 0 \\
0 & 0 & 2 & 2 & 0 \\
1 & 0 & 0 & 0 & 0 \\
0 & 1 & 0 & 0 & 0 \\
2 & 2 & 0 & 0 & 0 
\end{pmatrix}
\end{equation}
We impose the following condition:
\begin{equation}
\label{blockform} S_2 N_{2,1} S_2 =
\begin{pmatrix} \bullet & 0 & \begin{matrix} 0 & 0 & 0 \end{matrix} \\
0 & \bullet & \begin{matrix} 0 & 0 & 0 \end{matrix}\\
\begin{matrix} 0 \\ 0 \\ 0 \end{matrix}& \begin{matrix} 0 \\ 0 \\ 0 \end{matrix} & \setlength\fboxsep{0.6cm}\fbox{} \end{pmatrix}\;
,
\end{equation}
which leads to two restrictions for $S_2$:
\begin{equation*}
\left. \begin{aligned} (S_2 N_{2,1} S_2)_{51}&=& -2+2s_2+s_3+s_4 =
0 \\(S_2 N_{2,1} S_2)_{52}&=& -2-2s_2 = 0
\end{aligned}\right\rbrace \Rightarrow \left\lbrace \begin{aligned} s_3&=&4-s_4 \\ s_2 &=&
-1
\end{aligned}\right. \; .
\end{equation*}
The matrix element $s_4$ is left undetermined by this argument,
because with these two conditions also all other matrices
$N_{2,I}$ take the form as in eq. \eqref{blockform}, when they are
multiplied by $S_2$ from both sides. At this point our S-matrix
looks like
\begin{equation}
\label{Smats4}
\begin{pmatrix}
  \frac{1}{2} & \frac{1}{2} & 1 & 1 & 0 \\
  \frac{1}{2} & \frac{1}{2} & -1 & -1 & 0 \\
  \frac{1}{4} & -\frac{1}{4} & \frac{1}{2} & -\frac{1}{2} & 0 \\
  \frac{1}{4} & -\frac{1}{4} & -\frac{1}{2} & \frac{1}{2} & 0 \\
  1 &-1 &4-s_4 & s_4 & -1
\end{pmatrix} \; .
\end{equation}
The first two columns of the matrices $S_2$ and
$\mathtt{S}_{p,\alpha}$ are the same. The first two rows were
anyway the same from the beginning. This very much militates in
favour of a block diagonal $C_{2}(\alpha)$ apart from the good
reasons there are anyway because its smaller brother $C_{irr,2}$ (eq. \eqref{Cirrblock}) is also block diagonal. \\
We now take the matrix $K_2$ from equation \eqref{KExtDef} and our choice of the two matrix elements ${k^{(1)}}_{1}={k^{(1)}}_{2}=0$:
\begin{equation}
\label{Kgen}
 K_2=\begin{pmatrix}
   4 & 0 & 0 & 0 & 0 \\
   0 & -4 & 0 & 0 & 0 \\
   0 & 0 & 1 & \frac{1}{2} & 0\\
   0 & 0 & -1 & \frac{1}{2} & 0 \\
   0 & 0 & 0 & 0 & 1
 \end{pmatrix}\; ,
\end{equation}
In \protect\cite{Knuth:2006} we also shortly discuss the influence of this choice on $s_4$. \\
Together with the S-matrix in equation \eqref{Smats4}  we go through the calculations, which are needed to get to the coefficient matrices and look for a condition on $s_4$. The equations \eqref{PSK}, \eqref{defMp}
and \eqref{NPMP} lead us via the matrices $P_2$ and $M_{2,I}$ to
the matrices $N_{2,I}$. Here we are left with the argument that
the result should agree to the result of MF, which in turn
after some replacements agree with the result of M. Gaberdiel and H. Kausch.\\
We compare the following results for $N_{2,1}$ from the
calculations described:
\begin{equation*}
\begin{pmatrix}
 0 & 0 & 2 & 2 & 0 \\
 0 & 0 & 2 & 2 & 0 \\
 1 & 0 & 0 & 0 & 0 \\
 0 & 1 & 0 & 0 & 0 \\
 4-s_4 & s_4 & 0 & 0 & 0
\end{pmatrix}=\begin{pmatrix}
 0 & 0 & 2 & 2 & 0 \\
 0 & 0 & 2 & 2 & 0 \\
 1 & 0 & 0 & 0 & 0 \\
 0 & 1 & 0 & 0 & 0 \\
 2 & 2 & 0 & 0 & 0
\end{pmatrix}\; .
\end{equation*}
It follows that $s_4=2$, which gives also the other fusion coefficients correctly. This also gives some more ''symmetry''
to the S-matrix. The elements of the third and fourth column are
now the same modulo minus signs.\\
We have all matrices, which we
need for the extended block diagonalisation method for $p=2$. Still we need to find the matrix $C_2(\alpha)$, which both fulfils equation \eqref{C2def} and can be generalised to arbitrary $p$.

\subsection{Observations about Similar S-Matrices and the Matrix $\mathbf{C_2(\pmb{\alpha})}$}
We find a possible matrix $C_2$ fulfilling eq. \eqref{C2def} from the
matrices, which diagonalise $S_2$ and $\mathtt{S}_{2,\alpha}$,
to the same diagonal matrix $D_S$ (see eq. \eqref{DSDef})\; . If these
two diagonalising matrices are $U_1$ and $U_2(\alpha)$,
respectively, we have
\begin{equation}
\label{Sdiags}
 U_1^{-1} S_2 {U_1} = D_S = U_2(\alpha)^{-1} \mathtt{S}_{2,\alpha}
 {U_2}(\alpha)\; .
\end{equation}
This directly gives us a lot of possible matrices $C_2(\alpha)$ by rearrangement of equation \eqref{C2def}.
\begin{equation}
\label{CU1U2}
 C_{2}(\alpha)=U_1 {U_2}^{-1}(\alpha) \;,
\end{equation}
As the eigenspaces, we are looking at, are two or three
dimensional, there are quite a lot of matrices $U_1$ and
$U_2(\alpha)$ that meet our needs. Every possible basis of
eigenvectors spanning a particular eigenspace may be taken as the
columns of these matrices. In other words the columns may be any
linear combination of the eigenvectors in one eigenspace listed in 
tables \ref{appeigvec2} and \ref{appeigvec4} for the matrices $\mathtt{S}_{2,\alpha}$ and $S_2$ (with the matrix elements inserted, which we have found now), as long as they are linearly independent. \\
First we just state the one possible $C_2(\alpha)$ here, which
computes plainly using the listed eigenvectors as columns of $U_1$ and $U_2(\alpha)$ and get:
\begin{equation}
\label{C2part}
 C_{2,1}(\alpha)=\begin{pmatrix}
  \frac{9}{8} & \frac{7}{8} & - \frac{3}{2} + \frac{1}{2 \alpha} &
  \frac{1}{2 \alpha} & -\frac{1}{4 \alpha} \\
  -\frac{1}{8} & \frac{1}{8} & \frac{3}{2} - \frac{1}{2 \alpha} &
  -\frac{1}{2 \alpha} & \frac{1}{4 \alpha} \\
  0 & 0 & \frac{1}{2} - \frac{1}{\alpha} & \frac{1}{2} -
  \frac{1}{\alpha} & \frac{1}{2 \alpha} \\
  \frac{1}{8} & \frac{7}{8} & - \frac{1}{2 \alpha} &
  - \frac{1}{2} - \frac{1}{2 \alpha} & \frac{1}{4 \alpha} \\
  \frac{5}{8}  & \frac{11}{8} & -1 + \frac{1}{2 \alpha}  &
  \frac{1}{2} - \frac{1}{2 \alpha} & \frac{1}{4 \alpha}
 \end{pmatrix}\; .
\end{equation}
This does not fit our expectations.  This matrix does not have the
block structure, which our thoughts about the triples of
irreducible and indecomposable representations would suggest. We
also recall that we would like to have a matrix with the first
four rows equal to the matrix $C'_{2}(\alpha)$ from equation \eqref{Cprimealpha}. \\
But how much choice do we actually have for $C_2(\alpha)$? Or even better, what is the most general $C_2(\alpha)$, which we get from equation \eqref{CU1U2}, and are there others -- not in the form of
eq. \eqref{CU1U2} --, that fulfil equation \eqref{C2def}? The
answers are given by the following linear algebraic statement and
during its proof. \liketheorem{Lemma:}{} Let $S,\,\tilde{S} \in
M_{n \times n}(\mathbb{C})$ be two diagonalisable matrices, which
are diagonalised to the same matrix. Then they are similar to each
other and all matrices $C\in M_{n \times n}(\mathbb{C})$
fulfilling the equation
\begin{equation}
\label{CSCSlem}
 C S C^{-1}=\tilde{S} \; ,
\end{equation}
are given by the product of a particular $C=C_1$ times a matrix
$A\in M_{n \times n}(\mathbb{C})$, which commutes with $S$ or
$\tilde{S}$. Conversely any such product fulfils equation
\eqref{CSCSlem}. \qef \likeremark{Remark:}{} Equation
\eqref{CSCSlem} can also be defined with the matrices $S$ and
$\tilde{S}$ interchanged. But this does not make a difference,
when we go over from $C = C_1 A$ to $C^{-1}=A^{-1}{C_1}^{-1}$.
Note that the inverse of $A$
commutes with the same matrices as $A$ itself.\\
It is not needed here, but it is one line to see that any two
matrices $S$ and $\tilde{S}$, which are conjugate through a matrix
$C$ are diagonalised to the same diagonal matrix. If $S$ is
diagonalised by $P$,
\begin{equation*}
 P D P^{-1} = S = C \tilde{S} C^{-1} \; .
\end{equation*}
$\tilde{S}$ is diagonalised by $C^{-1}P$ to the same diagonal
matrix $D$. \qef \proof{$\!\!\!\!\!\!$:} We have already shown the
existence, because a particular solution for $C$ can be retrieved
via the eigenvectors of $S$ and $\tilde{S}$ from equation
\eqref{CU1U2}, as described
above. \\
Let $C$ and $C' $ be two matrices, which conjugate $S$ and
$\tilde{S}$ as in equation \eqref{CSCSlem}, so that we have
\begin{eqnarray}
\label{Cumg1}
 C \,S \,&=& \,\tilde{S} \,C \; ,\\
\label{Cumg2}
 C' \,S \,&=& \,\tilde{S} \,C'\; .
\end{eqnarray}
$A$ is defined by $A=C^{-1} C'$ and $C'=C A$ is plugged into the
last equation:
\begin{eqnarray*}
C\, A \, S \, &=& \, \tilde{S} \, C \, A\; .
\end{eqnarray*}
Now use equation \eqref{Cumg1} on the right hand side to get
\begin{eqnarray*}
C\, A \, S \, &=& \, C \, S \, A\; .
\end{eqnarray*}
Multiplying the inverse of $C$ shows that $A$ commutes with $S$.
$A$ commutes also with $\tilde{S}$ because its inverse does. This
is directly seen, when one goes through the analogous steps
starting
with $C= C' A^{-1}$ plugged into \eqref{Cumg1} and uses eq. \eqref{Cumg2}. \\
For the backwards direction we need only to multiply equation
\eqref{Cumg1} by an arbitrary matrix $A$, which commutes with $S$,
from the right side. We can interchange those two matrices on the
left hand side and find that $CA$ conjugates $S$ with
$\tilde{S}$.\qed We now need to find all matrices, which commute
with $\mathtt{S}_{2,\alpha}$. We continue to call them $A$ and
multiply the commutation relation of those two matrices by the
matrix ${U_2}(\alpha)$ and its inverse from opposite sides. We get
\begin{equation*}
{U_2}^{-1}(\alpha) \, A \, {U_2}(\alpha) \, {U_2}^{-1}(\alpha) \,
\mathtt{S}_{2,\alpha} \, {U_2}(\alpha) =  \, {U_2}^{-1}(\alpha)
\mathtt{S}_{2,\alpha} \, {U_2}(\alpha) \, {U_2}^{-1}(\alpha) \,
A \, {U_2}(\alpha)\; .
\end{equation*}
We simplify this with the help of equation \eqref{Sdiags}.
\begin{equation}
\label{DiagComm} {U_2}^{-1}(\alpha) \, A \, {U_2}(\alpha) \, D_S
\, = \, D_S \, {U_2}^{-1}(\alpha) \, A \, {U_2}(\alpha)\; .
\end{equation}
Hence we see that $A':={U_2}^{-1}(\alpha) \, A \, {U_2}(\alpha)$
has to commute with the diagonal matrix $D_S$ (eq. \eqref{DSDef}). All matrices $A'$ having this property are given by
\begin{equation*}
A'=\begin{pmatrix} (A')_{11} & (A')_{12} &(A')_{13} & 0 & 0 \\
(A')_{21} & (A')_{22} &(A')_{23} & 0 & 0 \\
(A')_{31} & (A')_{32} &(A')_{33} & 0 & 0 \\
0 & 0 & 0 & (A')_{44} & (A')_{45} \\
0 & 0 & 0 & (A')_{54} & (A')_{55}
\end{pmatrix}
\end{equation*}
with arbitrary $(A')_{ij}$ for $1 \leq i,j \leq 3$ or $4 \leq i,j
\leq 5$, so that the matrix has full rank.\\
Now we take this together with the definition of $A'$ beneath
equation \eqref{DiagComm} and the lemma to get via $A$ all
possible $C_{2,gen}(\alpha)$ (eq. \eqref{CU1U2}) from the one
particular $C_{2,1}(\alpha)$ (eq. \eqref{C2part}):
\begin{equation*}
 C_{2,gen}(\alpha)\,=\, C_{2,1}(\alpha) \, A \,= \, C_{2,1}(\alpha)
 \,{U_2}(\alpha) \, A' \, {U_2}^{-1}(\alpha)\; .
\end{equation*}
Of course, with so many unknowns the matrix $C_{2,gen}(\alpha)$
gets very lengthy. Now we simply require that the first four rows
of this matrix are equal to the matrix $C'_2(\alpha)$  from
equation \eqref{Cprimealpha}. We recall that this was justified by the
correspondence of $\tau$-dependent and $\alpha$-dependent
matrices described in section \ref{substaualpha}. We want to get an extension of Fuchs' approach, which
goes over to the latter one, when one projects to the irreducible
representations. In this case the matrix $C_2(\alpha)$ should
project to $C'_2(\alpha)$, which corresponds to $C_{irr,2}(\tau)$,
because the projection of $\mathtt{S}_{2,\alpha}$ corresponds to
$\mathsf{S}(2)$.\\
The fifth row then is the transpose of the following vector.
\begin{equation}
\label{5thline}
\begin{pmatrix}
  \frac{1}{2} - \frac{1}{8}(A')_{54} -
\frac{1}{8}(A')_{55}\\
-\frac{1}{2} + \frac{1}{8}(A')_{54}+ \frac{1}{8}(A')_{55}\\
1 + \left(\frac{1}{2} - \frac{1}{2\alpha} \right)(A')_{54} + \frac{1}{2\alpha}(A')_{55}\\
1 - \frac{1}{2\alpha}(A')_{54} + \left(\frac{1}{2} + \frac{1}{2\alpha}\right)(A')_{55}\\
\frac{1}{4\alpha}(A')_{54} -
\frac{1}{4\alpha}(A')_{55}
\end{pmatrix}\; .
\end{equation}
In analogy to $C_{irr,p}$ this matrix should be block diagonal.
This gives twice the same condition, which solves to
\begin{eqnarray*}
 (A')_{54}= 4 - (A')_{55}\; .
\end{eqnarray*}

 \subsection{Generalisation to Arbitrary Values of $\mathbf{p}$}
The last unknown, $(A')_{55}$, has been preliminary set to three because of
more aesthetic reasons. This way the matrix $C_2(\alpha)$
simplifies to
\begin{equation*}
   C_{2}(\alpha)=
 \begin{pmatrix}
  \begin{matrix}
  1 & 0 \\
  0 & 1
  \end{matrix} & 0   \\
   0 & \begin{matrix}
   \frac{3\alpha+2}{4 \alpha} & \frac{\alpha+2}{4 \alpha} & -\frac{1}{4 \alpha}\\
    \frac{\alpha-2}{4 \alpha} & \frac{3\alpha-2}{4
\alpha} & \frac{1}{4 \alpha}\\
    \frac{3\alpha+2}{2 \alpha} & \frac{5\alpha+2}{2
\alpha} & -\frac{1}{2 \alpha}
   \end{matrix}
 \end{pmatrix}\; .
\end{equation*}
It seems natural to have $(C_{2})_{55}=-1/(2\alpha)$. Firstly, it
fits to the grouping of terms, we have seen in section
\ref{substaualpha}. The factors of the $1/\alpha$-terms in the
last row are twice as large than in the first and second row. This
is expected because of the double multiplicities in the
indecomposable representation. Also the inverse of this matrix is
quite simple
\begin{equation}
 \label{C2expl}  C_{2}^{-1}(\alpha)=
 \begin{pmatrix}
   \begin{matrix}
  1 & 0 \\
  0 & 1
   \end{matrix} & 0   \\
   0 & \begin{matrix}
   2 & 1 & -\frac{1}{2}\\
    -1 & 0 & \frac{1}{2}\\
    \alpha+2 & 3\alpha+2 & -\alpha
   \end{matrix}
 \end{pmatrix}\; .
\end{equation}
This is very much in our favour, because we can now guess the
inverse of $C_{3}(\alpha)$ with not much effort. The last row in
every block is fixed looking at the result for $C_{3}(\alpha)$,
which it would lead to. We require once more that the first two
rows of both blocks are the blocks of the matrix $C'_{3}(\alpha)$.
We get
\begin{equation}
 \label{C3expl}  C_{3}^{-1}(\alpha)=
 \begin{pmatrix}
   \begin{matrix}
  1 & 0 \\
  0 & 1
   \end{matrix} & 0  & 0 \\
   0 &  \begin{matrix}
   2 & 1 & -\frac{1}{2}\\
    -1 & 0 & \frac{1}{2}\\
    \alpha+2 & 4\alpha+2 & -\frac{3}{2}\alpha
   \end{matrix} & 0\\
    0 & 0 & \begin{matrix}
   2 & 1 & -\frac{1}{2}\\
    -1 & 0 & \frac{1}{2}\\
    2\alpha+2 & 5\alpha+2 & -\frac{3}{2}\alpha
   \end{matrix}
 \end{pmatrix}\; .
\end{equation}
Its inverse is
\begin{equation*}
   C_{3}(\alpha)=
 \begin{pmatrix}
   \begin{matrix}
  1 & 0 \\
  0 & 1
   \end{matrix} & 0  & 0 \\
   0 &  \begin{matrix}
   \frac{2}{3}+\frac{1}{3\alpha} & \frac{1}{6}+\frac{1}{3\alpha} & -\frac{1}{6\alpha}\\
     \frac{1}{3}-\frac{1}{3\alpha} & \frac{5}{6}-\frac{1}{3\alpha} & \frac{1}{6\alpha}\\
   \frac{4}{3}+\frac{2}{3\alpha} &
\frac{7}{3}+\frac{2}{3\alpha} & -\frac{1}{3\alpha}
   \end{matrix} & 0\\
    0 & 0 & \begin{matrix}
   \frac{5}{6}+\frac{1}{3\alpha} & \frac{1}{3}+\frac{1}{3\alpha}  &  -\frac{1}{6\alpha}\\
    \frac{1}{6}-\frac{1}{3\alpha} &  \frac{2}{3}-\frac{1}{3\alpha} & \frac{1}{6\alpha}\\
  \frac{5}{3}+\frac{2}{3\alpha} &
\frac{8}{3}+\frac{2}{3\alpha} & -\frac{1}{3\alpha}
   \end{matrix}
 \end{pmatrix}\; .
\end{equation*}
This result, $C_{3}(\alpha)$, also determines the matrix $S_3$.
This matrix $S_3$ gives the correct fusion rules through our
extended block diagonalisation method.\\
Another choice of $(A')_{55}$, which we considered, is
$(A')_{55}=1$. This gives a very similar inverse of
$C_{2}(\alpha)$ and we can also guess the inverse of a potential
$C_{3}(\alpha)$, but this though similar has not the required same
elements as $C'_{3}(\alpha)$. Actually the rows of 
$C'_{3}(\alpha)$ are interchanged  and in these rows the first two
columns are exchanged in each block, while in the additional rows
there are also differences of one or the other minus sign.
Remarkably, this version gives a S-matrix and fusion rules, which
are qualitatively correct. Only the
multiplicities are wrong and turn out to be fractional.\\
With $C_2(\alpha)$ and $C_3(\alpha)$ we can guess the general matrix $C_p(\alpha)$, which is given in equation \eqref{Cpexpl}.

\bibliographystyle{halpha}
\bibliography{Bib}

\end{document}